\documentclass[aps,prb,preprint,groupedaddress,floatfix]{revtex4}
\usepackage{graphicx,latexsym,color}

\begin{document}


\title{Molecular dynamics simulations of reflection and adhesion behavior in Lennard-Jones cluster deposition}


\author{A. Awasthi and S. C. Hendy}
\affiliation{MacDiarmid Institute for Advanced Materials
and Nanotechnology, Industrial Research Ltd, Lower Hutt 5040, New Zealand}
\author{P. Zoontjens}
\affiliation{MacDiarmid Institute for Advanced Materials
and Nanotechnology, School of Chemical and Physical Sciences, Victoria University of Wellington,
Wellington 6140, New Zealand}
\author{S. A. Brown and F. Natali}
\affiliation{MacDiarmid Institute for Advanced Materials and Nanotechnology, Department of Physics and Astronomy,
University of Canterbury, Christchurch 8140, New Zealand}


\date{\today}

\begin{abstract}
We conduct molecular dynamics simulations of the collision of atomic clusters with a weakly-attractive surface. We focus on an 
intermediate regime, between soft-landing and fragmentation, where the cluster undergoes deformation on impact but remains largely intact, 
and will either adhere to the surface (and possibly slide), or be reflected. We find that the outcome of the collision is determined by the Weber number, {\it We} i.e. the ratio of the kinetic energy to the adhesion energy, with a transition between adhesion and reflection occurring as {\it We} passes through unity. We also identify two distinct collision regimes: in one regime the collision is largely elastic and deformation of the 
cluster is relatively small but in the second regime the deformation is large and the adhesion energy starts to depend on the kinetic energy. If the transition between these two regimes occurs at a similar kinetic energy to that of the transition between reflection and adhesion, then we find that the probability of adhesion for a cluster can be bimodal. In addition we investigate the effects of the angle of incidence on adhesion and reflection. Finally we compare our findings both with recent experimental results and with macroscopic theories of particle collisions.      
\end{abstract}


\maketitle



\section{Introduction}

The controlled self-assembly of atomic clusters is a very exciting approach to the construction of nanoscale electronic and photonic devices \cite{haberland93,meiwes00}. Some techniques exploit the tendencies of deposited atoms to aggregate into cluster-based nanostructures, such as the wire-like chains of clusters that can form on the naturally occurring step-edges on graphite \cite{howells02,francis96}. However many techniques involve the deposition onto a substrate of clusters that are first prepared in the gas phase. For this reason the collision of clusters with surfaces has been studied by many groups \cite{harbich00}. Once on the substrate, surface diffusion of clusters can lead to aggregation or pinning by surface defects \cite{kaiser02,howells02,francis96,brechignac02}. However, since the position of these surface defects or aggregates is usually random, the placement of the structures cannot be controlled. In contrast, by using lithographically defined V-grooves on passivated Si surfaces as templates \cite{partridge04}, a transition from adhesion to reflection can be exploited to form clean wire-like structures in the V-groove, if the deposition velocity is such that clusters bounce when they land away from the template, and stick (or slide) only in the V-groove.

Although the possibility of cluster reflection from hard surfaces was been postulated some time ago (see the ``phase" diagrams in Refs.~\cite{harbich00,hsieh92}), many early studies focused on other effects such as implantation \cite{yamamura94,Palmer02} or fragmentation \cite{pettersson93, chatelet92} at high energies or soft-landing at low energies \cite{cheng94}. At sufficiently high kinetic energies, one enters either the fragmentation regime where the cluster fragments \cite{pettersson93} or undergoes significant evaporation upon impact \cite{chatelet92}, or the implantation regime where the cluster buries itself in the surface \cite{Palmer02}. In the soft-landing regime, the incident kinetic energy is generally insufficient to overcome the adhesion between the surface and the cluster, resulting in collisions that always lead to adhesion. Only recently has an intermediate regime been identified where antimony and bismuth clusters were observed to undergo a transition from adhesion to reflection (while remaining substantially intact) as the kinetic energy was increased \cite{partridge04}. This transition has been now been exploited to assemble nanowires on a variety of patterned substrates \cite{reichel06}. 

While the study of collisions of solid bodies stretches back at least as far as Newton's treatment in the Principia, it is not clear to what extent the macropscopic theory of collisions applies to the collision of nanoscale bodies. An important quantity in macroscale collisions is Newton's coefficient of restitution, $e$, which is the ratio of the reflected to incident velocity, and which effectively measures the degree of inelasticity of a collision. Although the coefficient of restitution is often regarded as a material constant, it is known to depend on both the incident velocity and the degree of adhesion between the solid objects \cite{Johnson87}. Although inelastic collisions have been studied extensively for collisions of micro- or milliscale particles \cite{Hutchings77,Dave97,Wu03}, much less is known about the collision of nanoscale objects such as atomic clusters where adhesive forces will be much more important, and mechanisms for deformation will depend on size. At the macroscale, the weakly inelastic collision regime is frequently described using Hertzian contact mechanics \cite{Johnson87}. It is of interest to ask whether this theory applies at the nanoscale, and whether it can be used describe cluster deposition.  

Here we report on a detailed molecular dynamics study of the reflection of clusters with kinetic energies that lie between the soft-landing and fragmentation regimes, on surfaces with a weak attraction to the cluster. The resulting collisions span a range from weakly to strongly inelastic, resulting in little to substantial deformation of the cluster.  Previously, we reported on a reentrant adhesion transition that occurred in the strong deformation regime for normal collisions \cite{Awasthi06}. For a certain range of the cluster-surface interaction strength, we found the probability of adhesion was bimodal as a function of impact velocity. In this paper we will study the transition between adhesion and reflection for a much wider range of cluster-surface interaction strengths, including but not restricted, to the previous regime of interest where we observed the re-entrant transition. We will also consider collisions at non-normal angles of incidence, as these are obviously of interest for the cluster assembly process described in Ref~\cite{partridge04}. However, the main goal of this paper will be to make a comparison with macroscale theories of collision, and to consider previous experimental results \cite{partridge04,reichel06} in light of this comparison. 

We begin by discussing the methodology used in the simulations. We then discuss several individual collisions before looking at behavior averaged over many cluster orientations (in Ref.~\cite{Hendy06} we examined the effect of cluster orientation). We investigate the probability of reflection and adhesion as a function of the cluster size, the cluster-surface interaction strength and the angle of incidence. At the macroscale, the weakly inelastic collision regime is frequently described using Hertzian contact mechanics \cite{Johnson87}. We will compare our results to both this existing theory and recent finite element simulations of strongly plastic collisions at the milliscale \cite{Wu03,Wu05}. Finally, we consider the experimental data in Refs~\cite{partridge04,reichel06} in light of the understanding that we develop from the simulations. In particular, we attempt to estimate the velocity for the reflection-adhesion transition in the experimental systems.  

\section{Simulation Model and Methodology}

The interaction between atoms separated by a distance $r$ is modeled using a modified form of the Lennard-Jones potential \cite{barrat99}: 
\begin{equation}
\label{potential}
V(r) = 4\varepsilon \left[\left(\frac{\sigma}{r}\right)^{12} - C\left(\frac{\sigma}{r}\right)^6 \right] 
\end{equation}
for $r < r_c$, where $r_c$ is a cut-off chosen here to be $6 \sigma$ ($\sigma$ is the Lennard-Jones diameter and $\varepsilon$ is the depth of potential well). The large cut-off distance was chosen to ensure convergence both of the adhesion energy, and of the probability of reflection. For our system, the parameters $\varepsilon$ and $\sigma$ are the same for all atoms, although the constant $C$, which is applied as a scaling factor to the attractive part of the potential, is varied to control the attraction between surface and cluster atoms, while the atoms within the cluster and within the surface interact via the standard Lennard-Jones potential with $C=1.0$. Here we will consider collisions for values of $C$ between 0.2 and 0.7. Simulations using similar parameters have shown that varying $C$ between 0.5 to 1.0 leads to transitions from non-wetting to wetting behavior for a liquid droplet on a solid substrate \cite{barrat99}. In section~\ref{wettingsection}, we discuss the effect of $C$ on contact angle in our system.

We have simulated the collisions of various sizes of closed-shell Mackay icosahedra with a (111)-terminated fcc surface slab. The Mackay icosahedra are made up of 20 tetrahedrally shaped fcc units which share a common vertex. The (111)-terminated 7680-atom surface slab consists of a fixed bottom layer and 15 layers of dynamic atoms with dimensions of $11.7 \sigma \times 11.3 \sigma \times 10.3 \sigma$. Newtonian dynamics is applied to the central region of the slab while the outer region follows Langevin dynamics \cite{Langevin86} at a specified temperature $T$. The friction parameter in the Langevin thermostat is varied linearly from 0.0 at the Langevin-Newtonian interface to 2.0 at the boundary of the simulation cell \cite{Pomeroy02}. This block of 5846 Langevin atoms regulates the temperature of the 1344 Newtonian atoms and absorbs energy from the cluster impact. The surface computational cell is repeated periodically in the two dimensions parallel to the (111) surface plane, with no periodic boundary conditions applied in the z-direction. This arrangement of atoms was selected after carefully checking the convergence of the energetics and probability of reflection of the collisions. 

Both the Newton and Langevin equations of motion are integrated with a velocity Verlet scheme, with a time step $\delta t = 0.01 \tau$ where $\tau$ is the corresponding atomic time scale $\left(\frac{m\sigma^2}{\varepsilon}\right)^{1/2}$. If we were to use the Lennard-Jones parameters for Argon ($\sigma = 3.40$ \AA, $\epsilon = 1.65 \times 10^{-21}$ J) \cite{Kittel86}, our unit time corresponds to $\tau = 2.17$ ps. However, in the rest of this paper, unless noted otherwise all quantities are expressed in LJ reduced units using $\sigma$, $\varepsilon$ and $\tau$ as length, energy and time scales, respectively. 

The solid clusters and the surface slab were equilibrated at temperatures of $0.13 \frac{\varepsilon}{k_{B}}$ and $0.2 \frac{\varepsilon}{k_{B}}$, respectively for $10^4$ time steps to allow them to adopt relaxed configurations. The melting point of the  clusters was determined by constructing caloric curves \cite{Hendy05b} with the melting temperature of the 147-atom cluster found to be $0.33 \frac{\varepsilon}{k_{B}}$. Thus the simulations are conducted at temperatures well below the cluster melting point. At each size and velocity investigated we typically performed 50-100 impact simulations by varying the cluster orientation randomly.      

\section{Results}

In this section we present the results from a large number of simulated collisions for a range of parameters including the initial cluster velocity $v^*_0$ (in reduced units), the angle of incidence, the cluster-surface interaction $C$, and the cluster size. We begin by looking at the effect of $C$ on the wetting of the surface by the cluster. We then discuss several collisions in detail at fixed cluster orientation and cluster-surface adhesion strength. We then examine a large number of collisions averaged over cluster orientation for $C=0.35$ where we focus on the the reentrant adhesion transition. Next we examine the effect of varying $C$ between 0.2 and 0.7. For values of $C < 0.5$, we find a transition between adhesion and reflection that takes place at low velocities ($v^*_0 < 0.5$). For values of $C$ between 0.3 and 0.4 we observe a reentrant transition from reflection to adhesion at intermediate velocities ($0.5 < v^*_0 < 1.5$) followed once more by a transition to reflection. This reentrant transition occurs at the onset of a large deformation regime which increases the cluster contact area with the surface and thereby increases the adhesion energy. Finally, we look at collisions at non-normal incidence. We find that collisions at non-normal incidence follow a very similar behavior if one analyzes the results in terms of the normal velocity component.    

\subsection{\label{wettingsection}Relationship between $C$ and cluster contact angle}

\begin{figure}
\resizebox{\columnwidth}{!}{\includegraphics{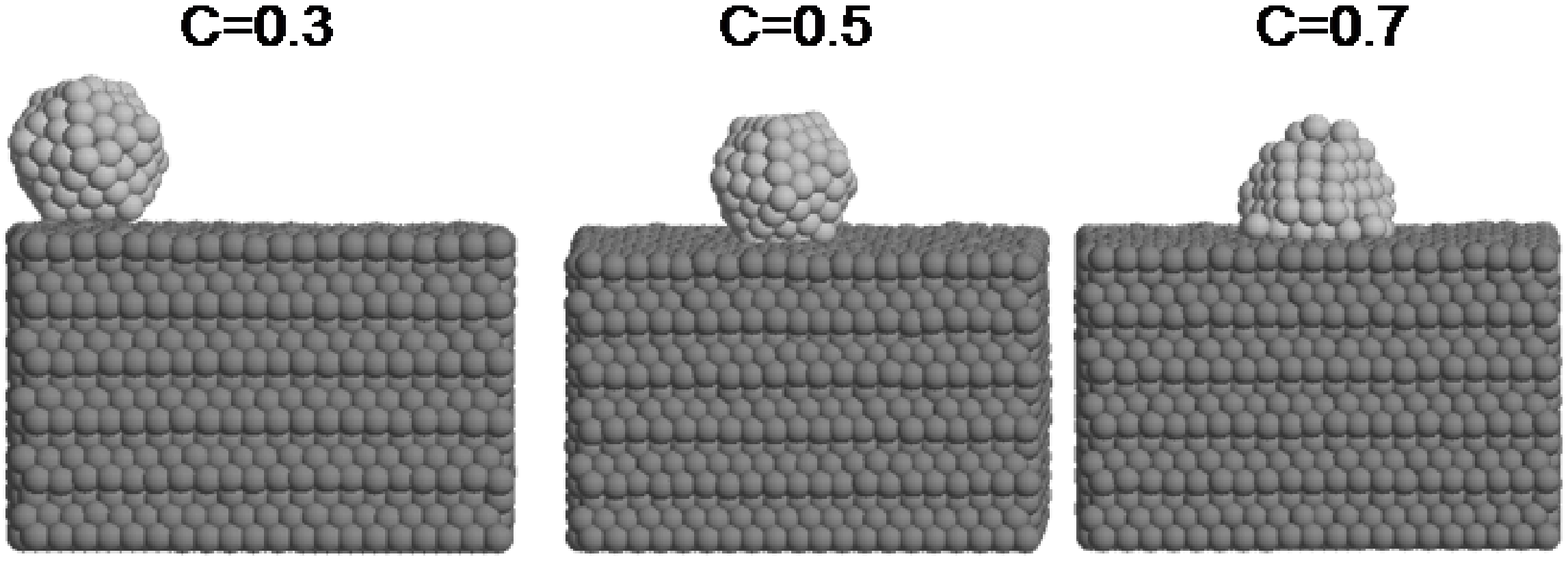}}
\caption{\label{contact}Snapshots showing a selection of 147-atom clusters after equilibration on
surfaces with various $C$ values. The solid clusters (top row) were equilibrated at $T^*=0.27$.}
\end{figure}

Figure~\ref{contact} shows a selection of snapshots of solid 147-atom clusters after equilibration on the (111) surface for $5 \times 10^5$ time steps for various values of $C$. Clearly, at $C=0.7$ the cluster can be regarded as "wetting" the surface. To estimate the contact angle we fitted a spherical cap to the positions of the cluster's surface atoms and at $C = 0.7$, the contact angle $\theta_w$ was found to be 88$^o$. We note that for $C < 0.5$, the solid clusters effectively don't wet the surface at all over the relaxation times examined here. Interestingly, although they remain bound, we observe significant diffusion of the cluster on the surface for $C < 0.5$ for both the liquid and solid clusters. 

In their experiments, Partridge et al \cite{partridge04} estimated contact angles of $\theta \sim 120^o$ for Sb and $30^o$ for Bi on SiO$_2$ using SEM imaging. However, as is possible that some of the clusters were molten prior to deposition, or even melted during the collision but later solidified as they cooled on the substrate, only tentative estimates of the the adhesion energies for Sb and Bi on SiO$_2$ can be made based on these contact angles. Furthermore, it is likely that the impact of the clusters lead to spreading of the cluster on the surface. Of course, the purpose of this study is not to precisely simulate the system in Ref~\cite{partridge04}, but rather to gain a general understanding of the reflection-adhesion transition. In Section~\ref{discuss}, we will consider the relationship between the simulated and experimental collisions in more detail.  

\subsection{A sample of individual collisions}

\begin{figure}
\resizebox{\columnwidth}{!}{\includegraphics{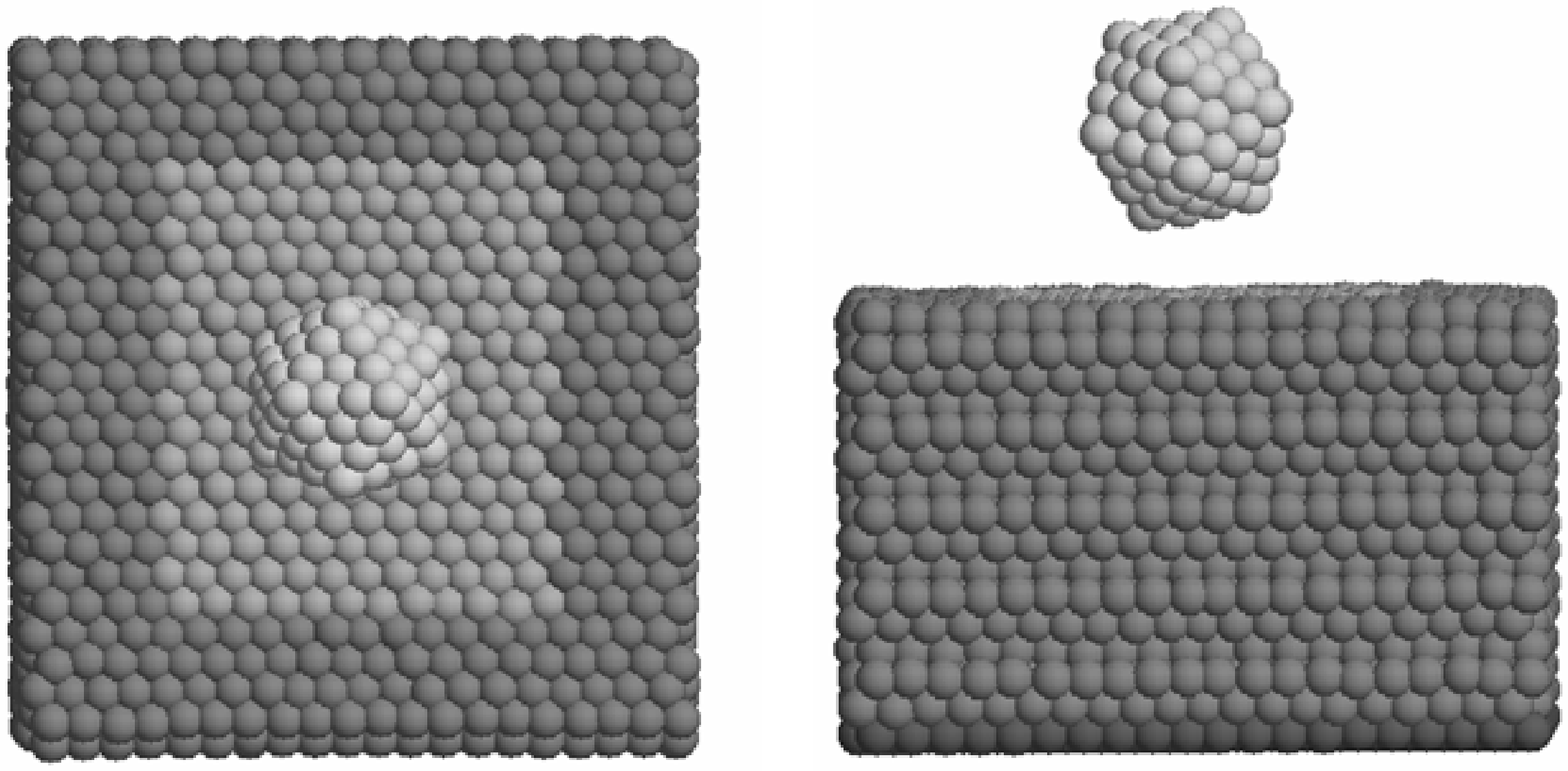}}
\resizebox{\columnwidth}{!}{\includegraphics{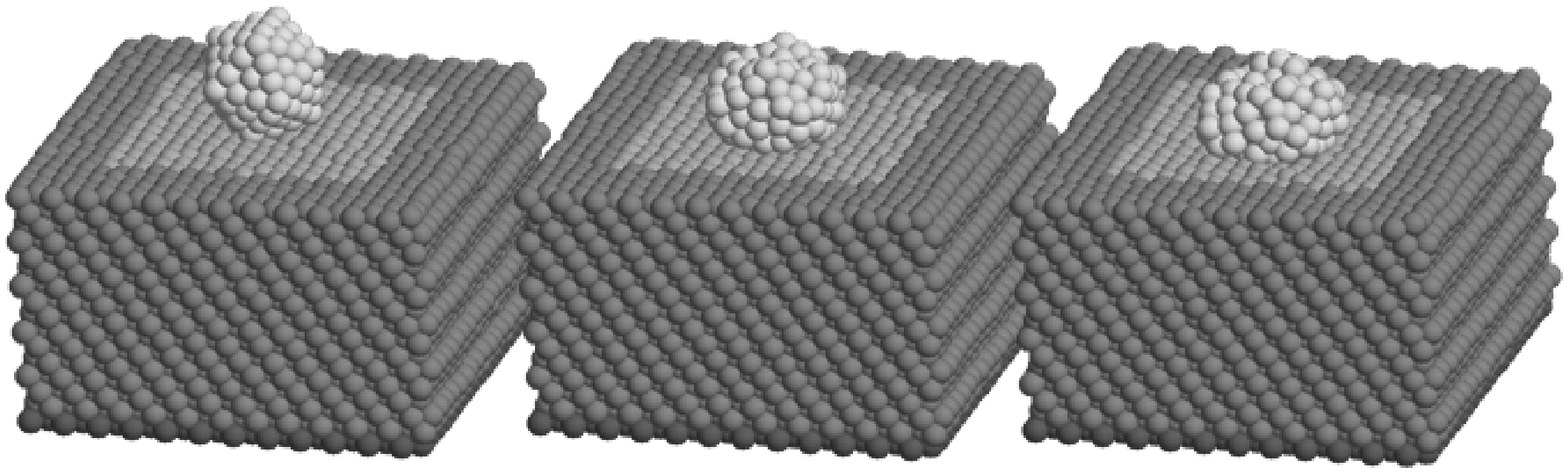}}
\caption{\label{ImpactSnap} The initial configuration for the collisions studied at $v^*_0 =$ 0.4, 1.6 and 2.6 (top). The lighter surface atoms follow Newtonian dynamics and the darker surface atoms follow Langevin dynamics. Snap shots of the 147 atom cluster colliding with the substrate with $C=0.35$ with initial velocities $v^*_0 = 0.4, 1.6$ and $2.6$ respectively at time $t^* = 9$ (bottom).}
\end{figure}

\begin{figure}
\resizebox{\columnwidth}{!}{\includegraphics{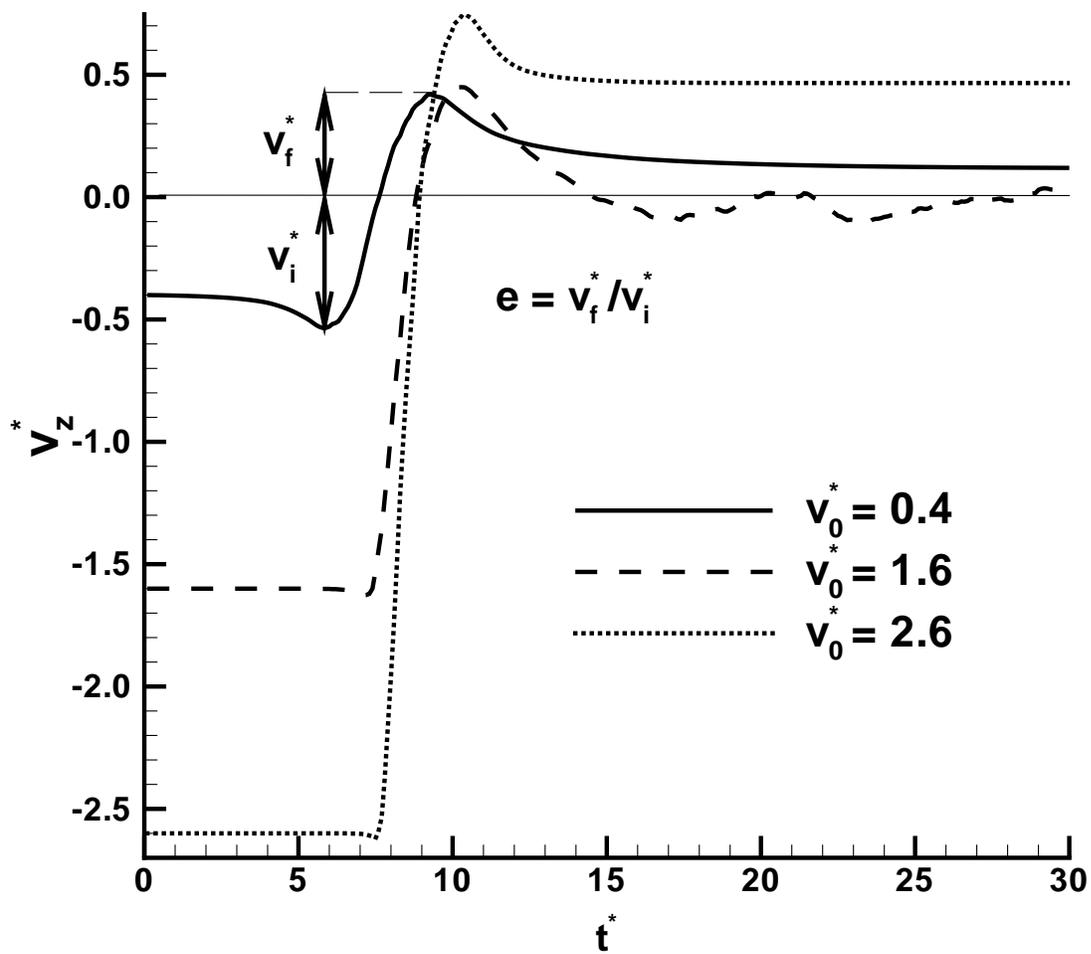}}
\caption{\label{vz} The z-component of velocity of the center of mass, $v_{z}^*$, of the 147 atom icosahedron plotted versus time for initial velocities $v^*_0 =$ 0.4, 1.6 and 2.6. The coefficient of restitution, $e$, is defined here as the ratio of the maximum velocity after the collision, $v_f^*$ to the maximum velocity before the collision, $v_i^*$.}
\end{figure}

Here we examine three collisions of the solid 147-atom icosahedral cluster, oriented edge-on to the surface as shown in Fig.~\ref{ImpactSnap}, at low, medium and high velocities with the cluster-surface interaction strength fixed at $C=0.35$. This will allow us to introduce the quantities we will later use to analyze a larger number of collisions as parameters are varied. Fig.~\ref{vz} plots the evolution of the velocities of the cluster center of mass of center for clusters with initial velocities $v^*_0 = 0.4, 1.6$ and $2.6$. Interestingly, we observe that the clusters with intial velocities of $v^*_0 = 0.4$ and $2.6$ escape, while the cluster that had initial velocity $v^*_0=1.6$, is bound and its velocity oscillates about zero. We will see in subsequent sections that this propensity for clusters to stick at intermediate velocities is typical for $C=0.35$. Note the slight acceleration and deceleration of the cluster in each case just before and just after the collision due to the attraction between the cluster and the surface. For this reason we define the coefficient of restitution for the collision, $e$, as the ratio of the peak velocity after collision to that before the collision as shown in the figure i.e. $e = - v^*_f/v^*_i$. This is a convenient choice for the definition for $e$ in the presence of a significant adhesive interaction between the cluster and the surface. A more conventional choice would define $e$ to be the ratio of the initial and final velocities but this leads to $e=0$ for clusters that adhere to the surface. In our simulations, where clusters frequently adhere to the surface, our definition of $e$ it provides a measure of the reflected kinetic energy `available' for escape. In what follows we refer to quantities evaluated at the moment of peak velocity after collision (the `pull-off') with a subscript $f$. 

\begin{figure}
\resizebox{\columnwidth}{!}{\includegraphics{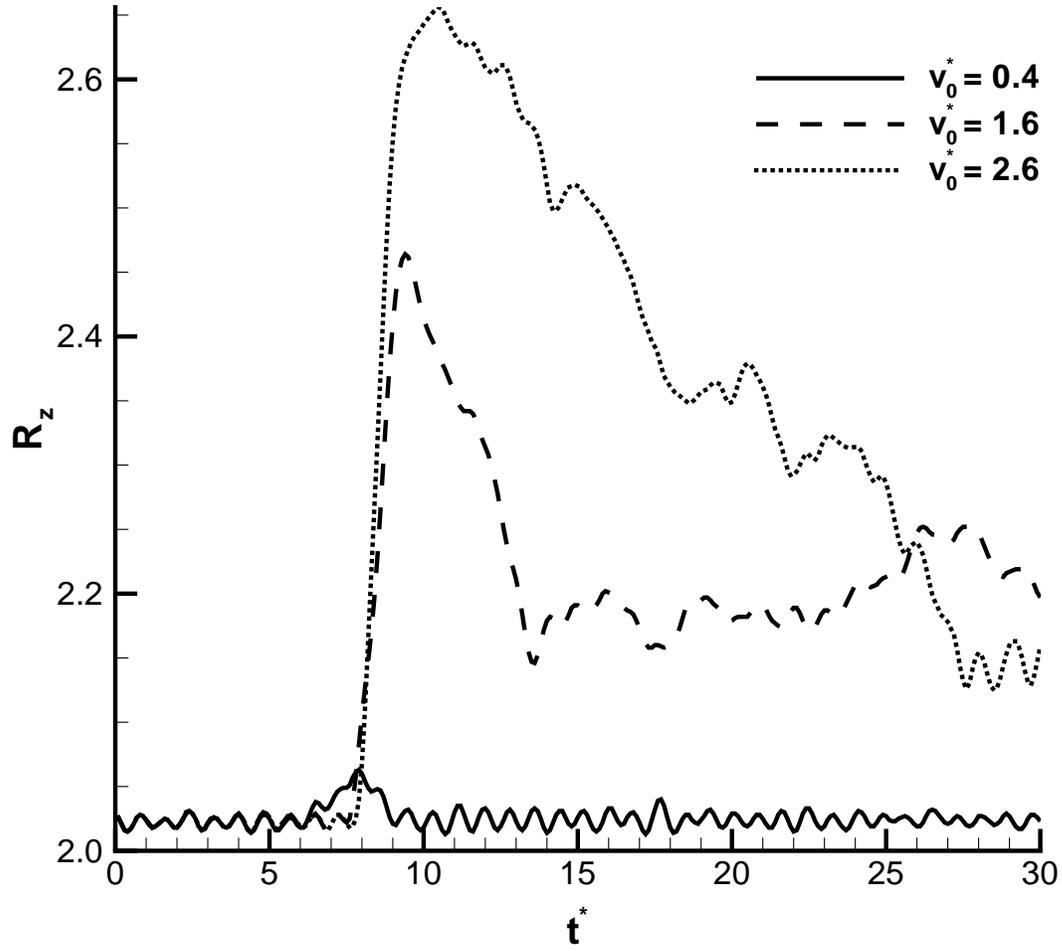}}
\caption{\label{Rz} The evolution of the z-component of the radius of gyration, $R_{z}$, of 147 atom cluster colliding with the substrate with $C=0.35$ for initial velocities, $v^*_0 =$ 0.4, 1.6 and 2.6.}
\end{figure}

To examine the deformation of the cluster during the collision (illustrated in Fig~\ref{ImpactSnap} by snapshots of the collisions taken at $t^* = 9$, which is close to the moment of maximum deformation) we use the radius of gyration, $R^g=(R_x^2+R_y^2+R_z^2)^{1/2}/\sqrt{2}$, where for example, $R_z$ is defined as
\begin{equation}
R_z = \left( \frac{1}{N}\sum^N_i (x_i^*-x_{cm}^*)^2 + (y_i^*-y_{cm}^*)^2 \right)^{1/2}.
\end{equation} 
Figure~\ref{Rz} shows the evolution of $R_z$ for the three collisions. In all three cases $R_z$ increases sharply at the collision. At $v^*_0=0.4$ the deformation is small and reversible suggesting that the deformation is largely elastic. Both of the higher velocity collisions show evidence of irreversible (plastic) deformation. For instance at $v^*_0=1.6$, while some of the initial deformation relaxes, there is strong permanent (plastic) deformation. For $v^*_0=2.6$, the cluster bounces and the radius of gyration increases by about 30 \% but eventually settles down to a value below that of the cluster that adhered to the surface at $v^*_0=1.6$.  

\begin{figure}
(a){\includegraphics[height=7.5cm]{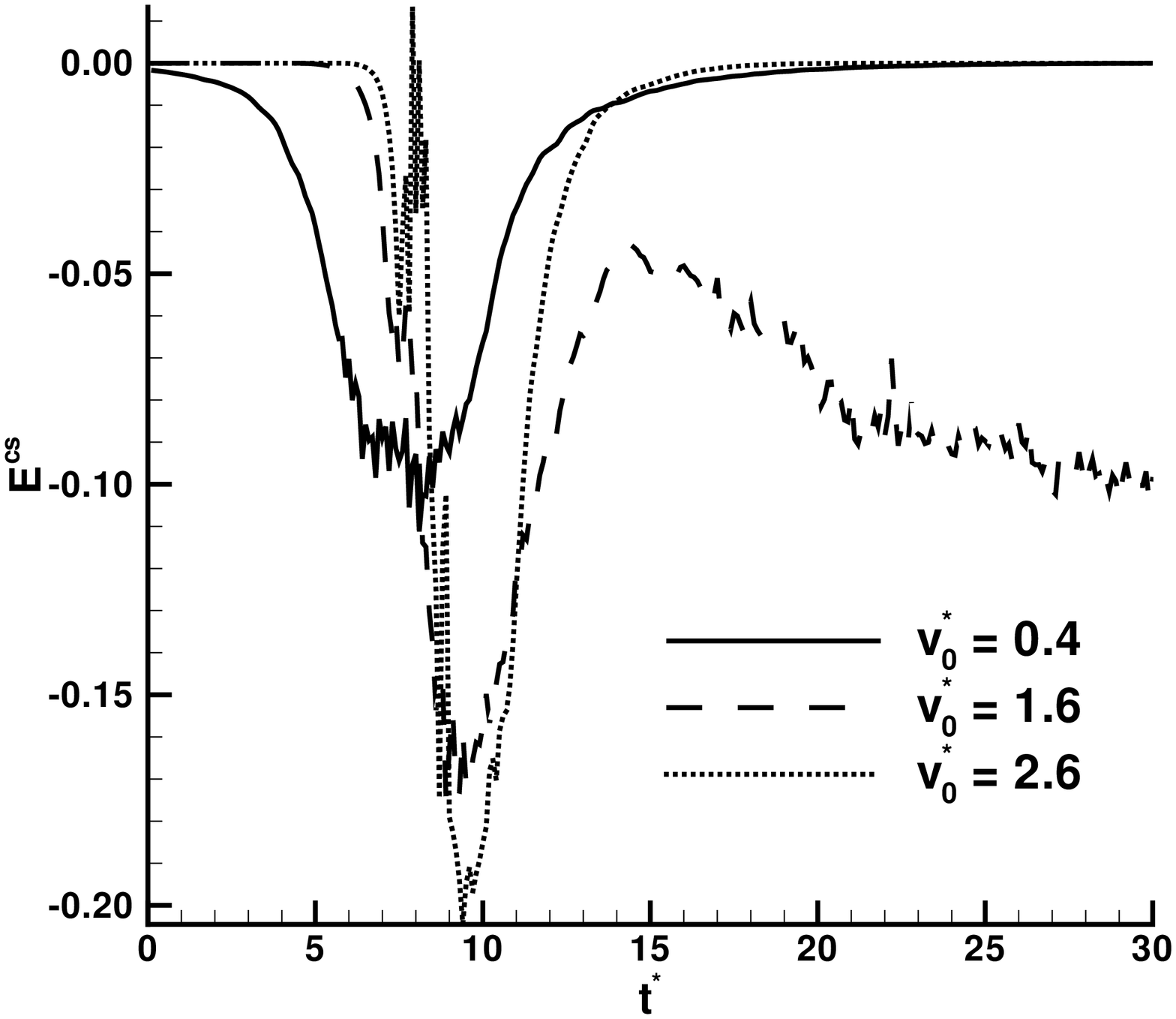}}
(b){\includegraphics[height=7.5cm]{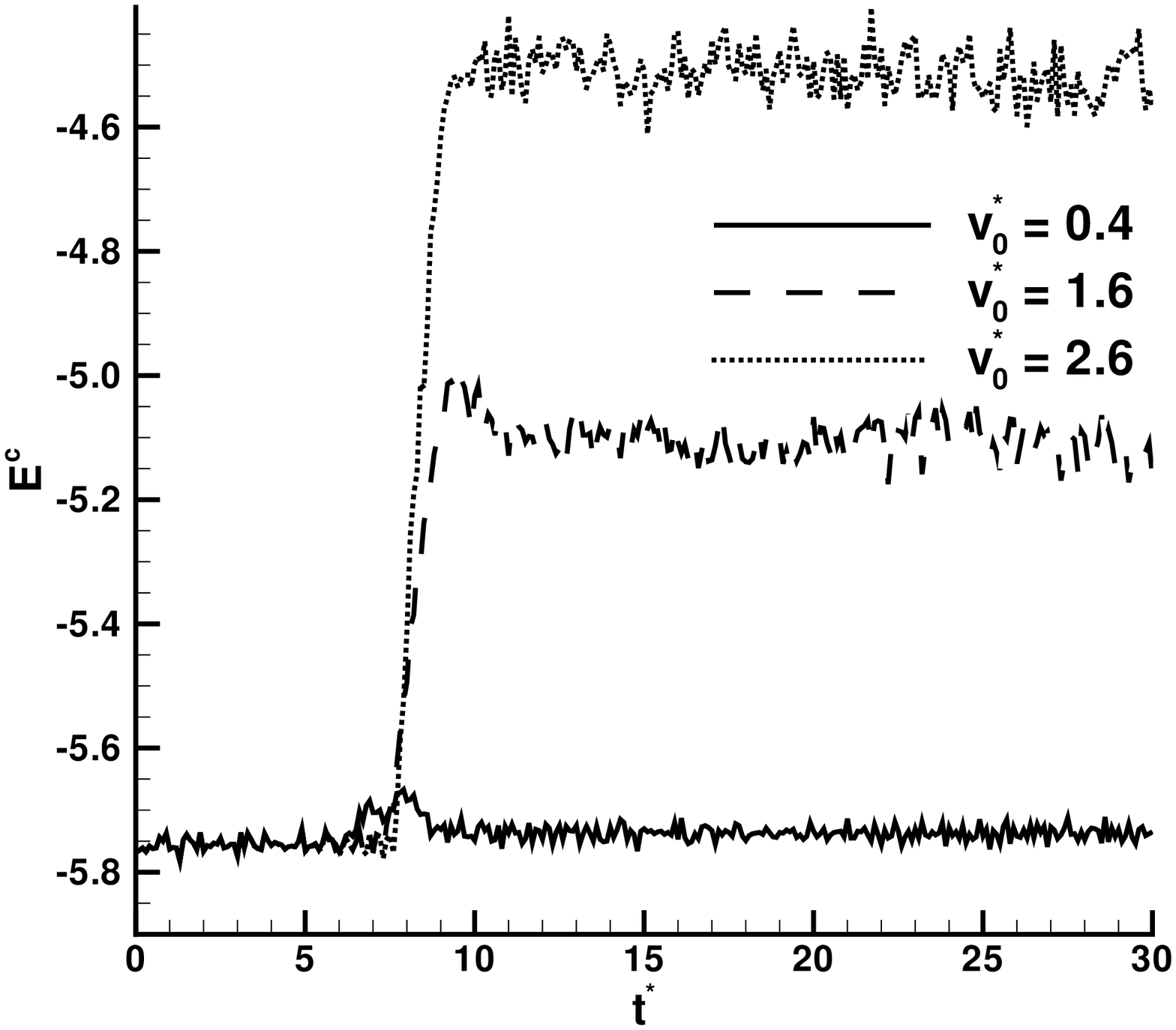}}
(c){\includegraphics[height=7.5cm]{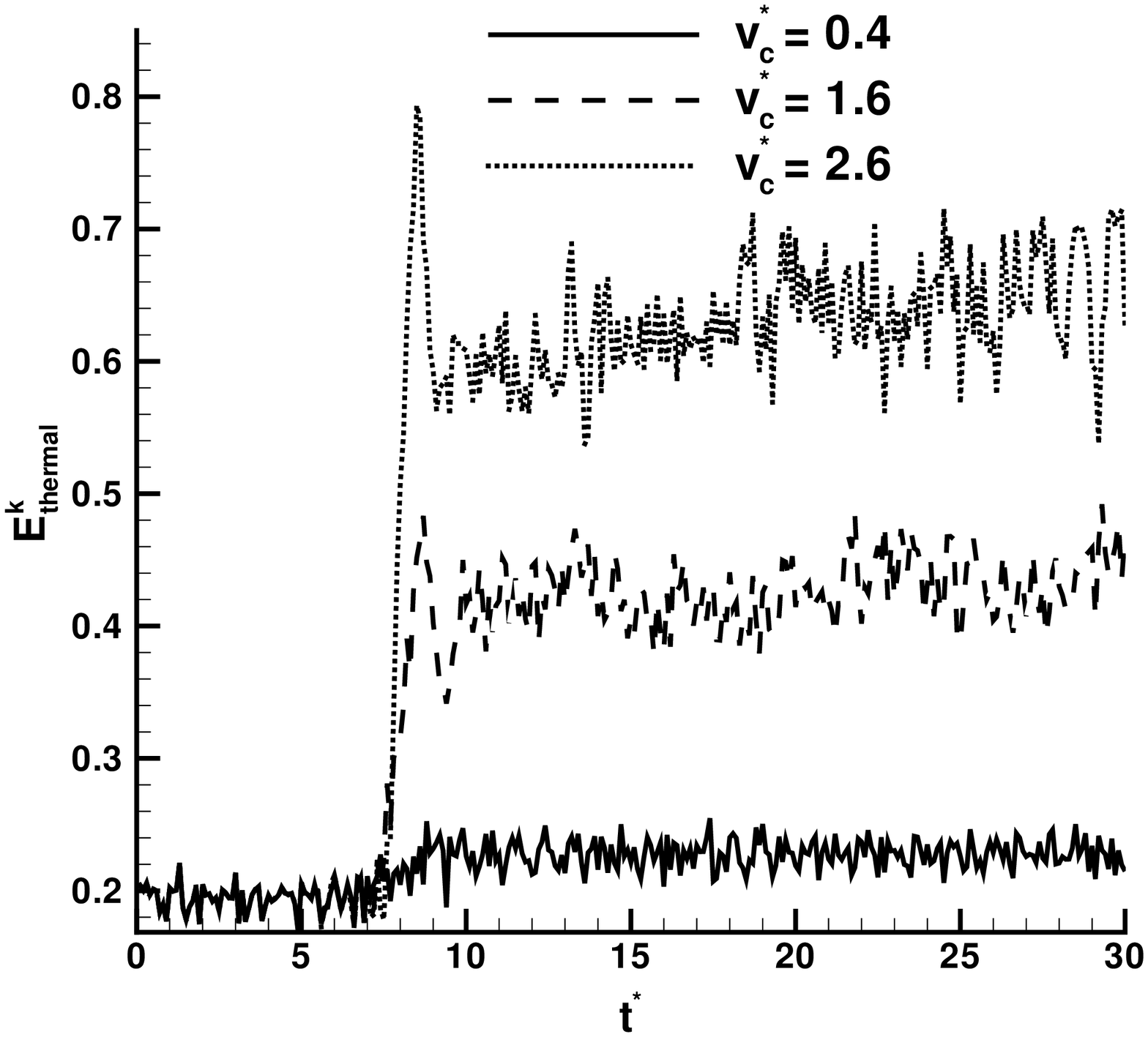}}
\caption{\label{PotE} The evolution of (a) the cluster-surface potential energy, $E^{cs}$, (b) the cluster potential energy, $E^{c}$, and (c) the thermal kinetic energy, $E^k_{thermal}$ of 147 atom cluster (equilibrated initially at $T^*=0.13$) colliding with the substrate with $C=0.35$ for initial velocities, $v^*_0 =$ 0.4, 1.6 and 2.6.}
\end{figure}

The total cluster potential energy per atom, $E^{pot}$, is the sum of cluster internal energy per atom, $E^{c}$, and cluster-surface interaction energy per atom, $E^{cs}$ (the adhesion energy can be defined as $E^a=-E^{cs}$). As seen in Fig.~\ref{PotE}(a), $E^{cs}$ at first decreases (i.e. the adhesive energy increases) as the cluster approaches the surface due to the attraction between the cluster and surface. During the collision $E^{c}$ increases as the cluster is deformed by the impact. At $v^*_0=0.4$, $E^{c}$ returns to its precollision value (Fig.~\ref{PotE}(b)) confirming that the collision is elastic (as was indicated by the deformation in Fig.~\ref{Rz}). Likewise, for the two more energetic collisions the change in $E^{c}$ is permanent, confirming that the collision is largely plastic. In the case of the collision at $v^*_0$=1.6 which results in adhesion, we see that there is a subsequent relaxation of $E^{cs}$ as the cluster begins to equilibrate with the surface. We note that the plastic deformation of the clusters during impact leads to a corresponding increase in cluster temperature as shown in Fig.~\ref{PotE}(c). Further at both $v^*_0=1.6$ and $2.6$ the clusters reach temperatures above the free cluster melting point ($T_c^*$ = 0.33). An inspection of snapshots of the clusters after impact strongly suggest that indeed the clusters have melted.

Finally, the ratios of the kinetic energy to the adhesion energy of the clusters at the pull-off (which is the point of zero net centre of mass force), $E^K_f/E^{a}_f$, are 1.04, 0.67 and 1.68 for the $v^*_0 = 0.4, 1.6$ and $2.6$ collisions respectively. As expected if $E^K_f/E^{a}_f < 1$ the cluster will adhere to the surface whereas if $E^K_f/E^{a}_f >1$ the cluster will be reflected. We note that for a spherical droplet, this ratio is equivalent to the Weber number, {\it We}, often used in fluid mechanics. In the rest of this paper, we will define the Weber number to be {\it We} $ = E^K_f/E^{a}_f$ following Ref.~\cite{Quere} where it was used to analyse the bouncing of liquid droplets. 

\subsection{\label{sec:pstick}Probability of adhesion averaged over cluster orientation for $C=0.35$}

Here the collision of the 147, 309 and 561 atom icosahedral clusters with (111)-terminated surface slab is examined at a fixed cluster-surface interaction strength of $C=0.35$ and for initial cluster velocities between 0.2 and 3.2 incident at 90$^o$ to the surface. For each cluster size and velocity, 50-100 trials were performed but between each trial the cluster was randomly reorientated prior to the collision. The effects of cluster orientation are reported in detail in Ref.~\cite{Hendy06}. There we identified three characteristic orientations that lead to distinct collision behavior: vertex-first, edge-first and facet-first. At low velocities, the collision depends strongly on cluster orientation with vertex-first collisions more likely to lead to reflection and edge-first collisions more likely to result in adhesion. At high velocities, the collisions depended only weakly on orientation. In what follows we average out the effect of cluster orientation on the collision probability. 

\begin{figure}
\resizebox{\columnwidth}{!}{\includegraphics{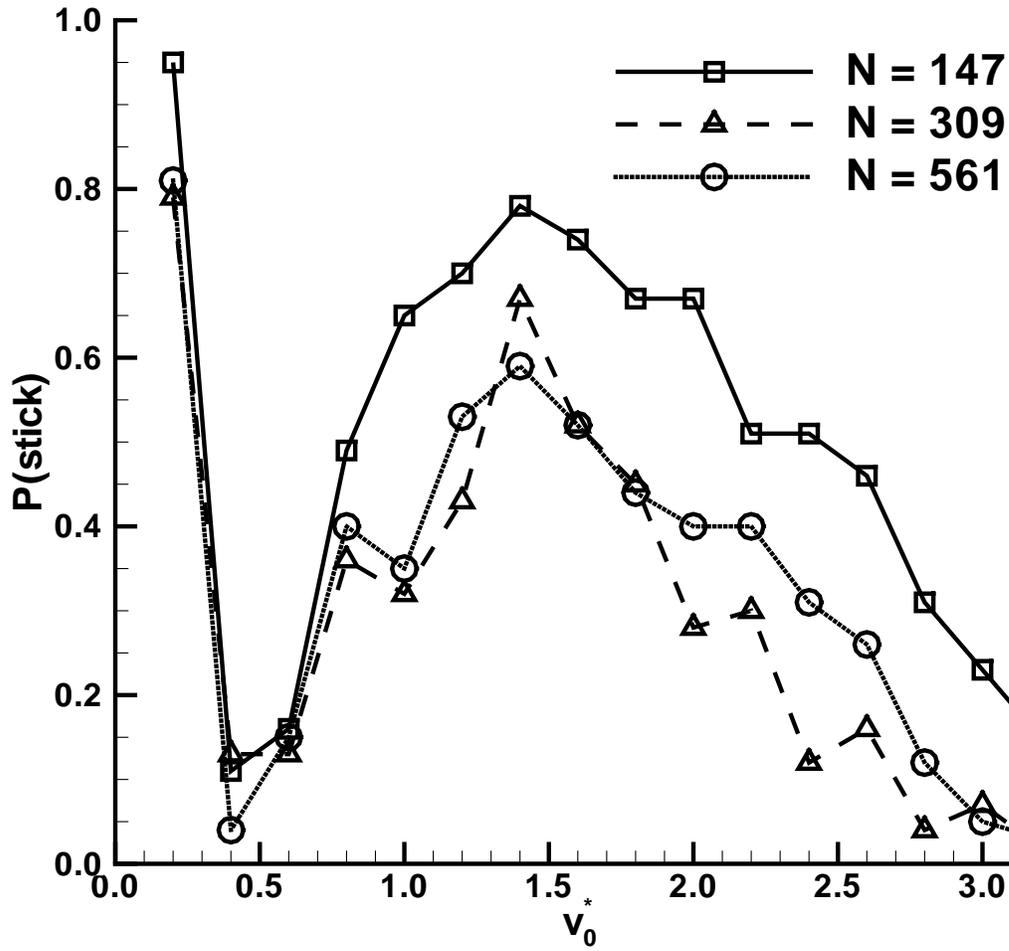}}
\caption{\label{StickFigure} The probability of adhesion versus the initial velocity, $v^*_{0}$), averaged over 100 trials with random orientations with $C=0.35$ for the three different cluster sizes incident on the flat surface.} 
\end{figure}

Fig.~\ref{StickFigure} shows that the probability of adhesion as a function of incident velocity is bimodal for the solid clusters of each size (as reported in Ref.~\cite{Awasthi06}). Clusters adhere to the surface at very low velocities ($v^*_0 < 0.3$) but start bouncing at intermediate velocities $0.3 > v^*_{0} > 0.5$. At higher velocities, $v^*_{0} > 0.5$, the adhesion probability increases but once more starts to decrease for $v^*_{0} > 1.5$. As we will show in more detail below, this is because there are essentially two deformation regimes. For $v^*_0 < 0.3-0.4$, little deformation occurs, so that the area of contact (and hence the adhesion energy) depends only weakly on the incident velocity. At higher velocities i.e. for $v^*_{0} > 0.5$, the deformation starts to grow substantially, leading to an increase in contact area (and adhesion energy) which depends strongly on velocity. In this strong deformation regime ($v^*_0 > 0.5$), the adhesion energy initially dominates the reflected kinetic energy as the deformation produces a larger contact area. This is evidenced by the increase in adhesion probability between $0.5 < v^*_0 < 1.5$. Eventually, the reflected kinetic energy begins to dominate adhesion ($v^*_0 > 1.5$) and the probability of adhesion decreases. All three cluster sizes display this bimodality, although the larger clusters are less likely to adhere to the substrate in general. 

The deformation can be studied by examining how the change in the radius of gyration depends on velocity. Figure~\ref{deformFigure} shows the relative change in radius of gyration, $\Delta R_f^g / R^g$, at the moment of peak reflected velocity as a function of incident velocity. There are clearly two regimes: at small velocities, $v^*_0 < 0.5$, $\Delta R_f^g / R^g$ is effectively zero, while for $v^*_0 > 0.5$ this relative deformation grows quadratically with $v^*_0$. Note that the relative deformation is only weakly dependent on cluster size. 

\begin{figure}
(a){\includegraphics[height=7.5cm]{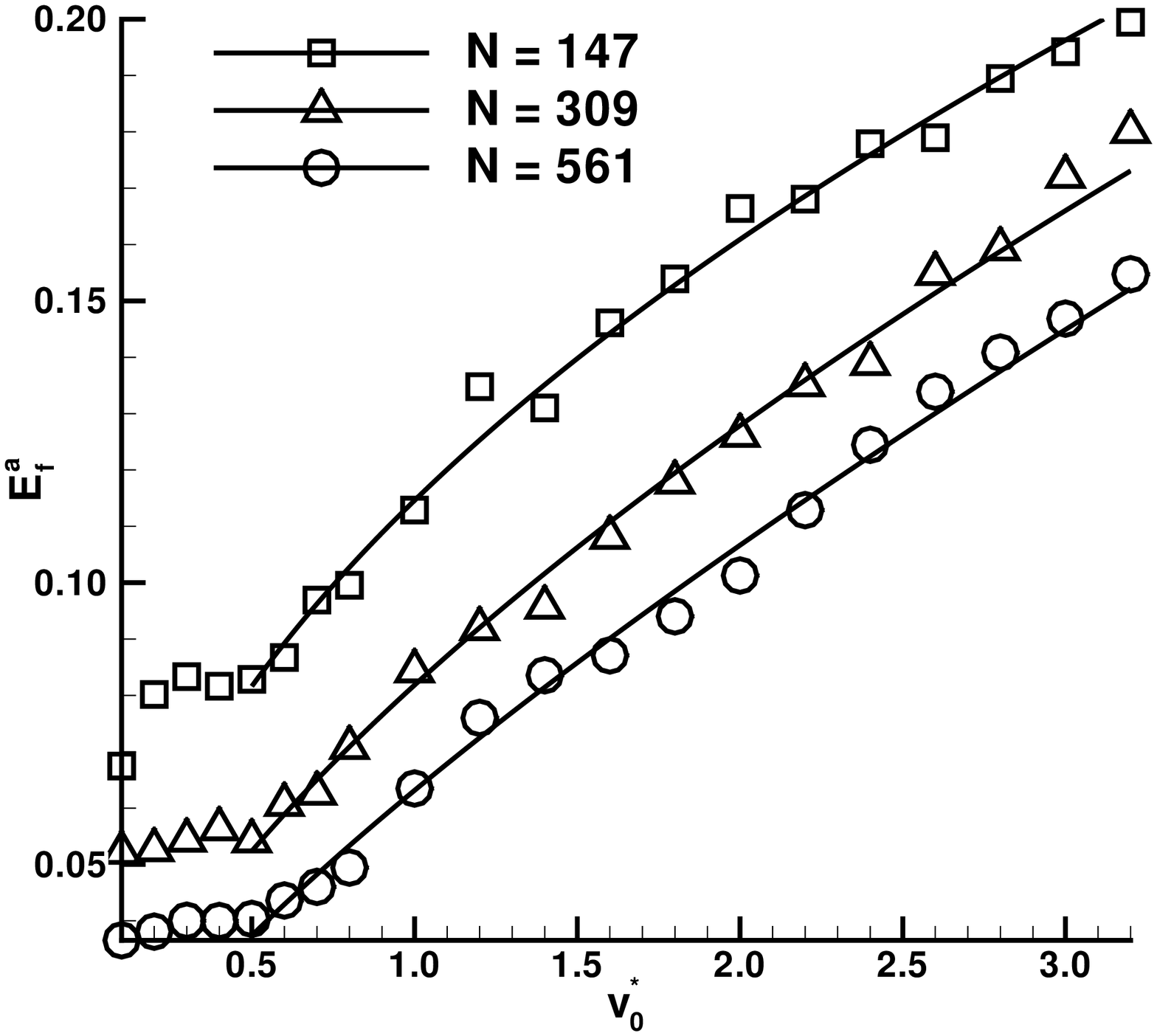}}
(b){\includegraphics[height=7.5cm]{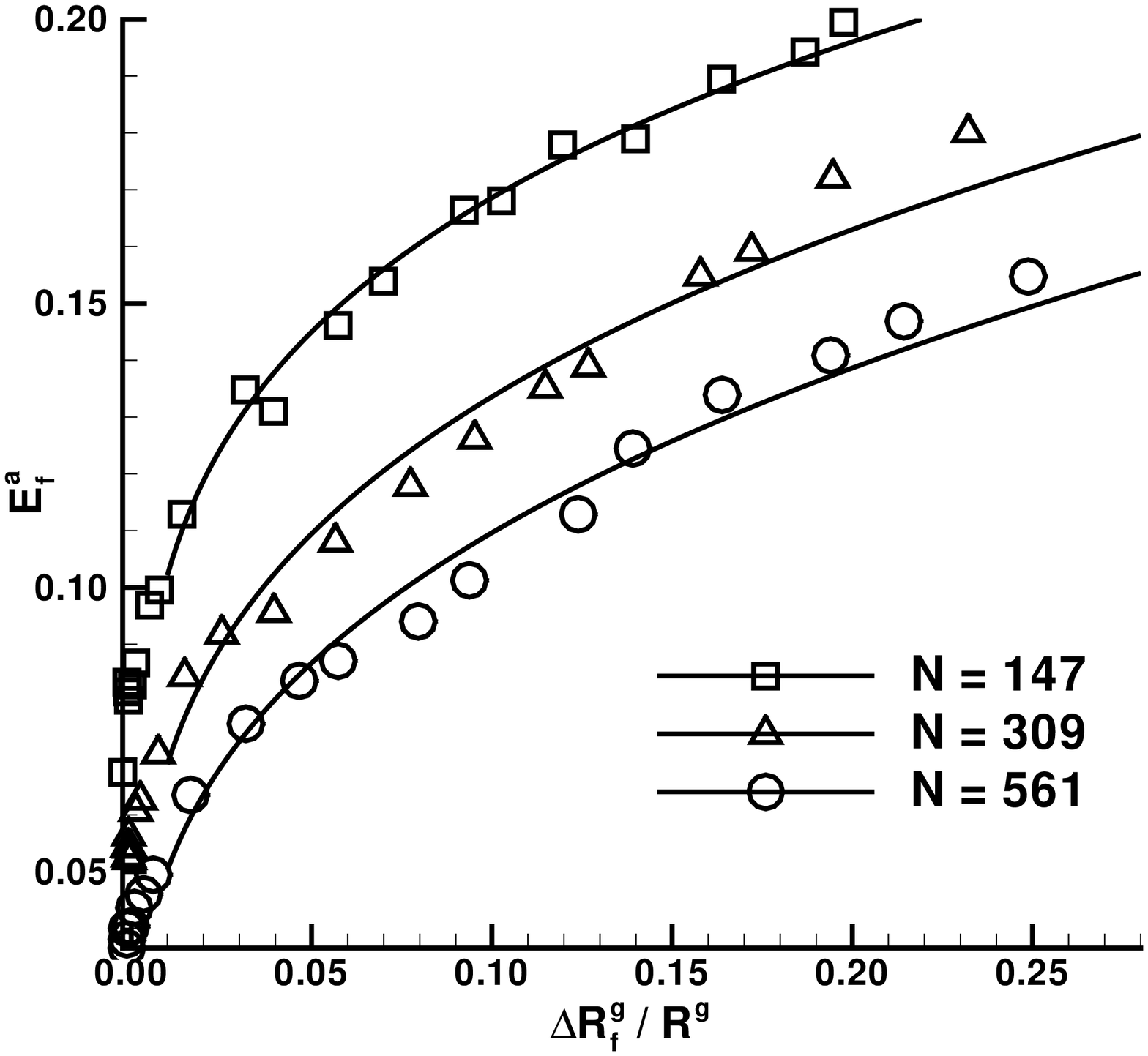}}
(c){\includegraphics[height=7.5cm]{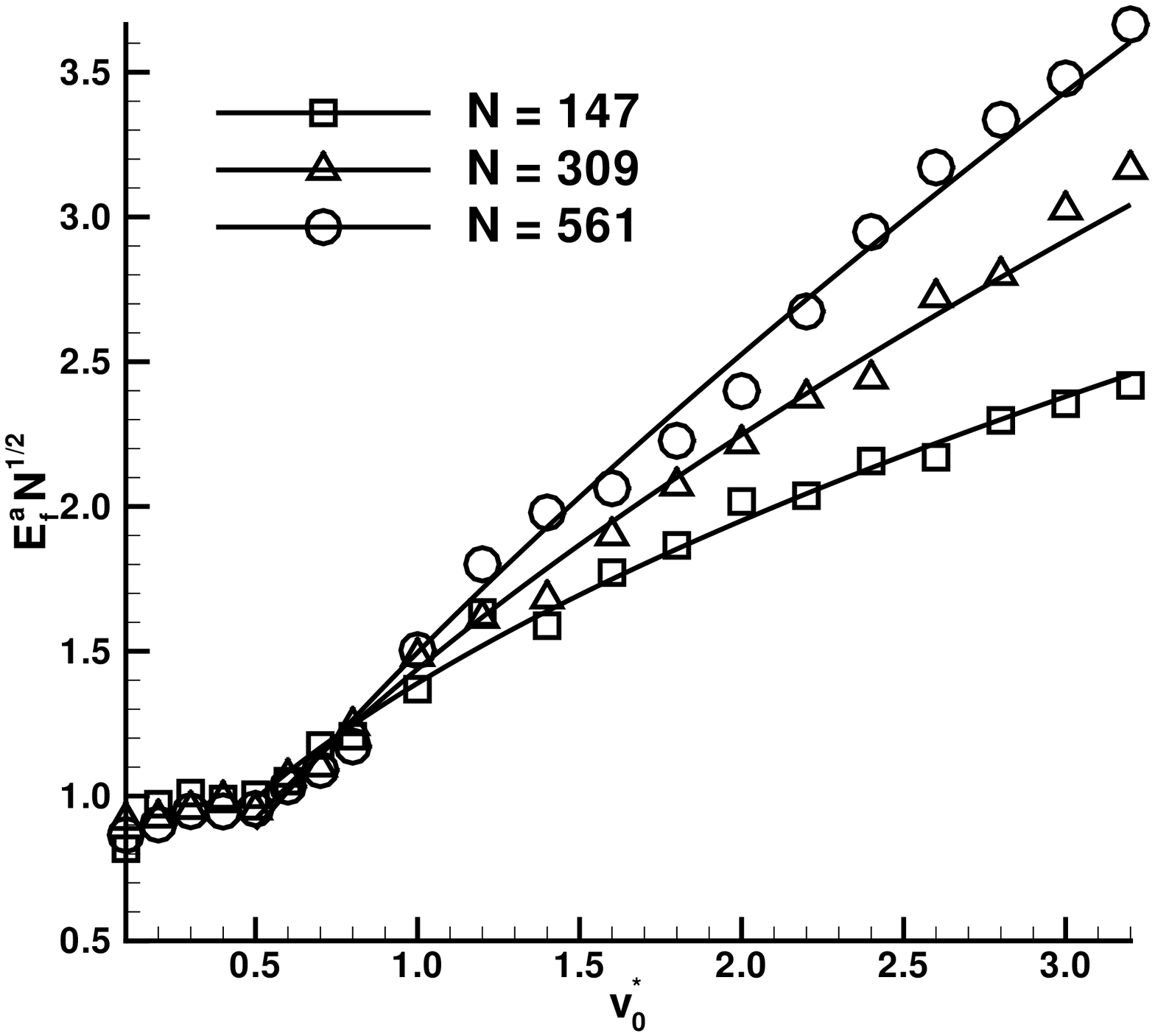}}
\caption{\label{EavsV}For $C=0.35$ (a) the adhesion energy per atom, $E^a_f$ at `pull-off' as a function of impact velocity, (b) $E^a_f$ as a function of $\Delta R_f^g / R^g$, and (c) $E^a_f N^{1/2}$ as a function of impact velocity for the three different cluster sizes. From c) we observe that the total adhesion energy ($N \times E^a_f$) scales as $N^{1/2}$ in the small deformation regime.} 
\end{figure}

The deformation at the pull-off also affects the adhesion energy. Fig.~\ref{EavsV}(a) shows the adhesion energy per atom of the cluster at the moment of peak reflected velocity, $E^{a}_f=-E^{cs}_f$, versus impact velocity. $E^{a}_f$ is roughly constant at low velocities, $v^*_0 < 0.5$ where, as noted above, there is little deformation during impact. However, for velocities $v^*_0 > 0.5$, where there is strong deformation, $E^{a}_f \sim \left(v^*_0\right)^{0.5 - 0.7}$ as shown in Fig.~\ref{EavsV}(a). Figure~\ref{EavsV}(b) shows the dependence of $E^{a}_f$ on the relative deformation $(\Delta R_f^g / R^g )$ of the cluster at the moment of peak reflected velocity. In Fig.~\ref{EavsV}(c), $E^{a}_f N^{1/2}$ is plotted as a function of velocity for all three cluster sizes to illustrate the dependence of $E^{a}_f$ on $N$. From this figure we note that at low velocities ($v_0^* < 0.5$), in the elastic regime, $E^{a}_f \sim N^{-1/2}$ as is shown by the coincidence of the values of $E^{a}_f N^{1/2}$ for all cluster sizes. For $v_0^* \gg 0.5$, in the strong deformation regime, the dependence on $N$ is weaker and scales more like $N^{-1/3}$. This is consistent with a `pancaking' of the cluster on the surface (see the snapshots in Fig.~\ref{ImpactSnap}).

\begin{figure}
\resizebox{\columnwidth}{!}{\includegraphics{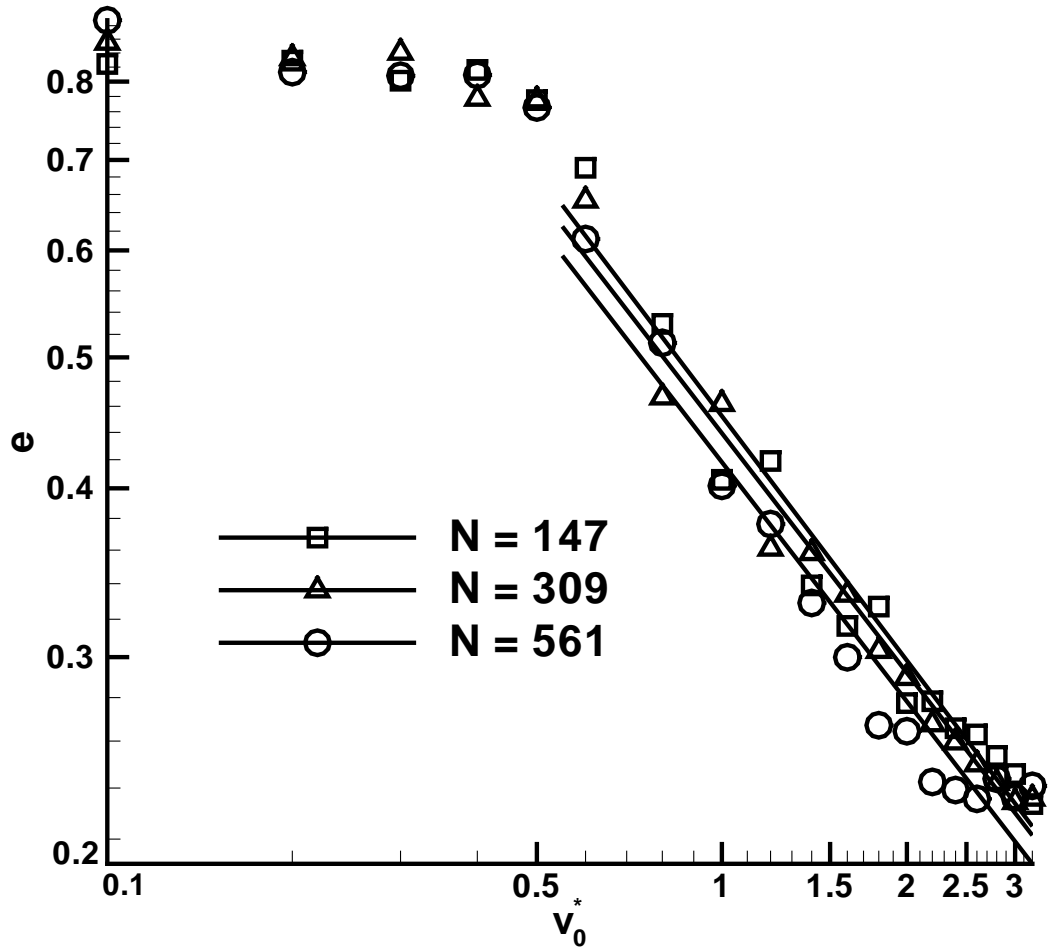}}
\caption{\label{corFigure} The coefficient of restitution, $e$, for the three different cluster sizes on the flat surface as a function of initial velocity at $C=0.35$ for normal incidence.}   
\end{figure}

Figure~\ref{corFigure} shows the variation of coefficient of restitution, $e$, with the impact velocity. Each data point shown in figure represents an average over 100 trials for each cluster size. The data shows a rough trend for $e$ to decrease as the cluster size increases. $e$ is approximately constant for low velocities but shows a strong dependence on velocity at  $v^*_0 > 0.5$. In this strong deformation regime, the dependence of $e$ on velocity varies as $e \sim \left(v^*_0\right)^{-0.6}$ for the 147, 309 and 561-atom icosahedra. This dependence on velocity is much stronger than that predicted by small deformation contact mechanics \cite{Johnson87} which predicts a $\left(v^*_0\right)^{-0.25}$ decay in the coefficient of restitution.
However it is close to the $(v_0^*/v^\dagger)^{-0.5}$ dependence of $e$ found by finite-element simulations of strongly plastic collisions \cite{Wu03}.

\begin{figure}
\resizebox{\columnwidth}{!}{\includegraphics{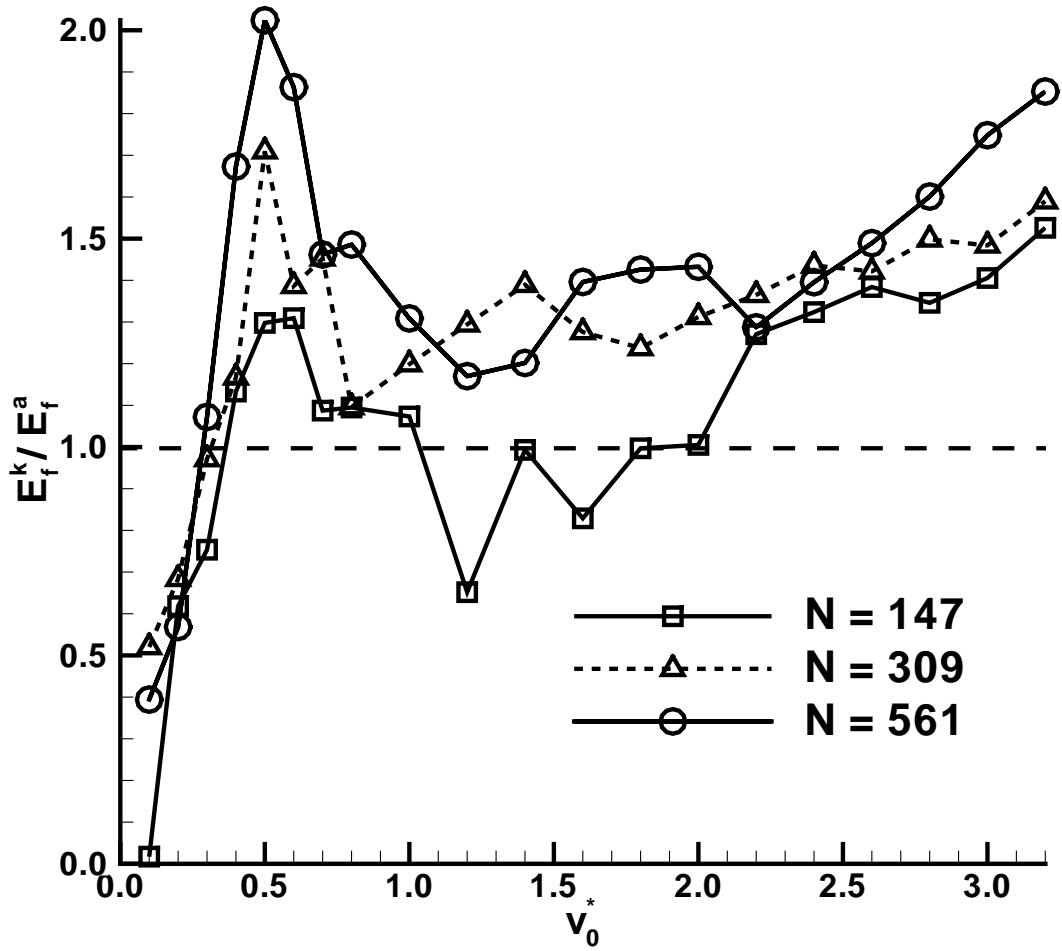}}
\caption{\label{weber} The ratio of the reflected kinetic energy $E^K_f$ to $E^{a}_f$ or Weber number {\it We} at the point of peak reflected velocity for the different size clusters with $C=0.35$.} 
\end{figure}

Thus, we have clearly identified two collision regimes by looking at the cluster at the pull-off point: a low-deformation regime, with a constant coefficient of restitution and adhesion energy, and a strong deformation regime where both $e$ and $E^a_f$ depend strongly on the initial velocity. To understand how this affects the probability of adhesion, we have plotted the Weber number, {\it We}, at the moment of peak reflected velocity
in Fig.~\ref{weber}. We note that {\it We} correlates well with the probability of reflection. As the average value of {\it We} approaches and then exceeds 1, the number of clusters being reflected dramatically increases. From this plot, we can identify several collision regimes for the solid clusters. At low velocities ($v^*_0 < 0.5$) the clusters undergo little deformation, and both the adhesion energy and coefficient of restitution are approximately constant. In this low deformation regime the reflected kinetic energy grows to dominate the adhesion, and consequently the probability of adhesion decreases with impact velocity. However for $v^*_0 > 0.5$, we begin to see substantial deformation of the cluster where both the coefficient of restitution and the adhesion energy depend on the velocity. Initially, the adhesion energy dominates, leading to an increase in adhesion probability. However, at high velocities the reflected kinetic energy begins to dominate again and the probability of adhesion decreases once more. 
 
\subsection{The effect of varying $C$}

Figure~\ref{c_Stick} illustrates the effects of varying the strength of cluster-surface attraction, $C$, showing the adhesion probability of 147-atom icosahedron as a function of the impact velocity. It is seen that the adhesion probability strongly depends on $C$ and the transition from adhesion to reflection of the cluster is observed as the value of $C$ is decreased from $C$=0.7 to $C$=0.2. The bimodal behavior of adhesion probability is evident at $C=0.3-0.4$ but disappears outside this range as either the reflected kinetic energy dominates at small $C$ or the adhesion energy dominates at large $C$. We note that the difference in reflection probability for surfaces with different adhesion energies has been exploited for device fabrication \cite{reichel06}.

\subsection{The effect of the angle of incidence}

\begin{figure}
\resizebox{\columnwidth}{!}{\includegraphics{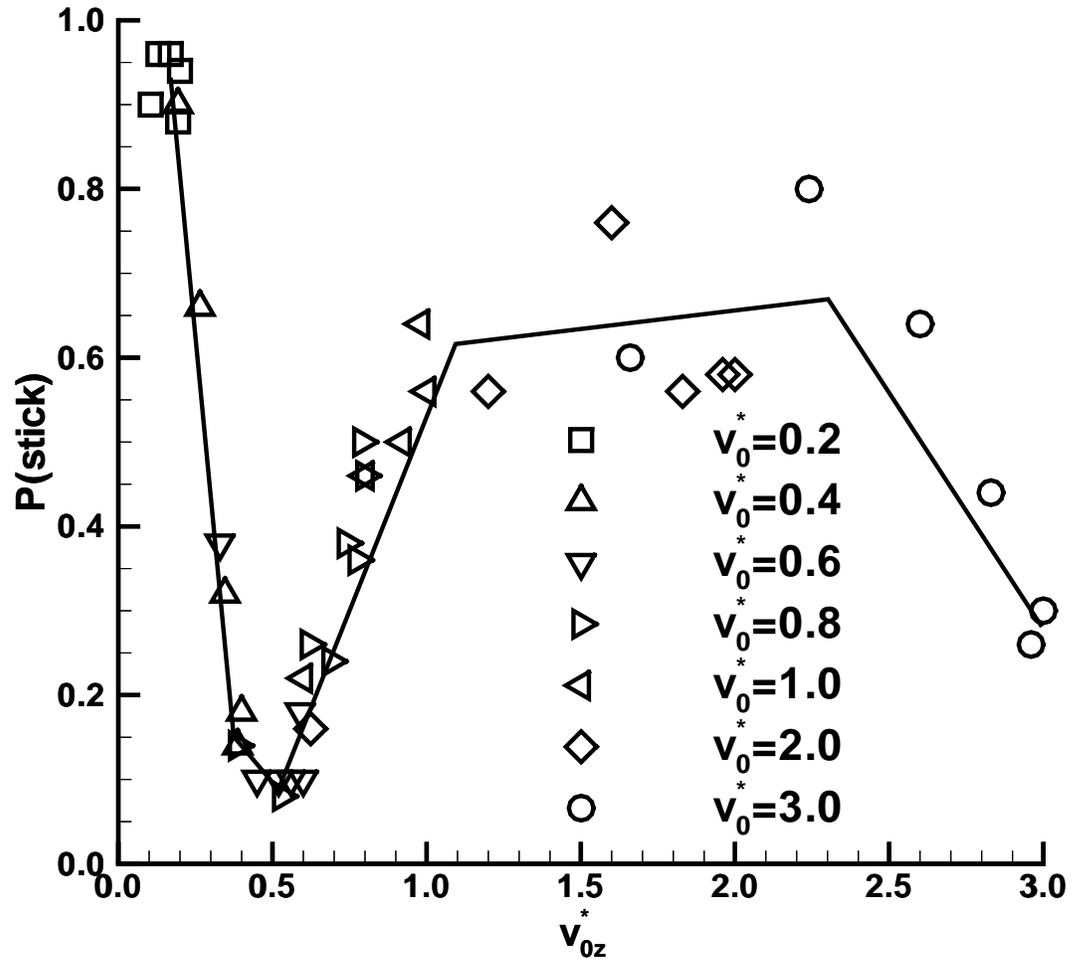}}
\caption{\label{ObliqueStickFigure} The probability of adhesion versus the normal component of the initial velocity, $v^*_{0z}$, averaged over 50 trials with $C=0.35$ for non-normal incidence of 147-atom icosahedron on the substrate. The lines shown are to guide the eye.} 
\end{figure}

Finally, we have considered the adhesion of the 147-atom cluster at non-normal angles of incidence for $C=0.35$. In Fig.~\ref{ObliqueStickFigure} we have plotted the probability of adhesion versus the normal velocity component ($v^*_{0z}$) at angles from 30-90 degrees (again the cluster was randomly reorientated between each trial). Note that the probability of adhesion as a function of the normal velocity component for the non-normal collisions is very similar to that of the normal collisions in Fig.~\ref{StickFigure}. Thus the probability of adhesion is largely determined by the normal component of the incident velocity and hence for the oblique collisions, we define the coefficient of restitution as the ratio of normal velocity components after and before impact: $e = - v^*_{fz}/v^*_{iz}$. With this definition, we found that the magnitude and dependence on velocity of the coefficient of restitution in the oblique case was very similar to that seen in the normal incidence case. Indeed, this is consistent with analysis of the angular momentum of the reflected clusters, which shows that they do not gain significant amounts of rotational kinetic energy after the collisions (typically < 1\% of the translational kinetic energy) even after impacts at highly oblique angles. However, we would expect the conversion of translational to rotational kinetic energy during collisions to be more important at larger cluster sizes. 

\begin{figure}
{\includegraphics[height=7.5cm]{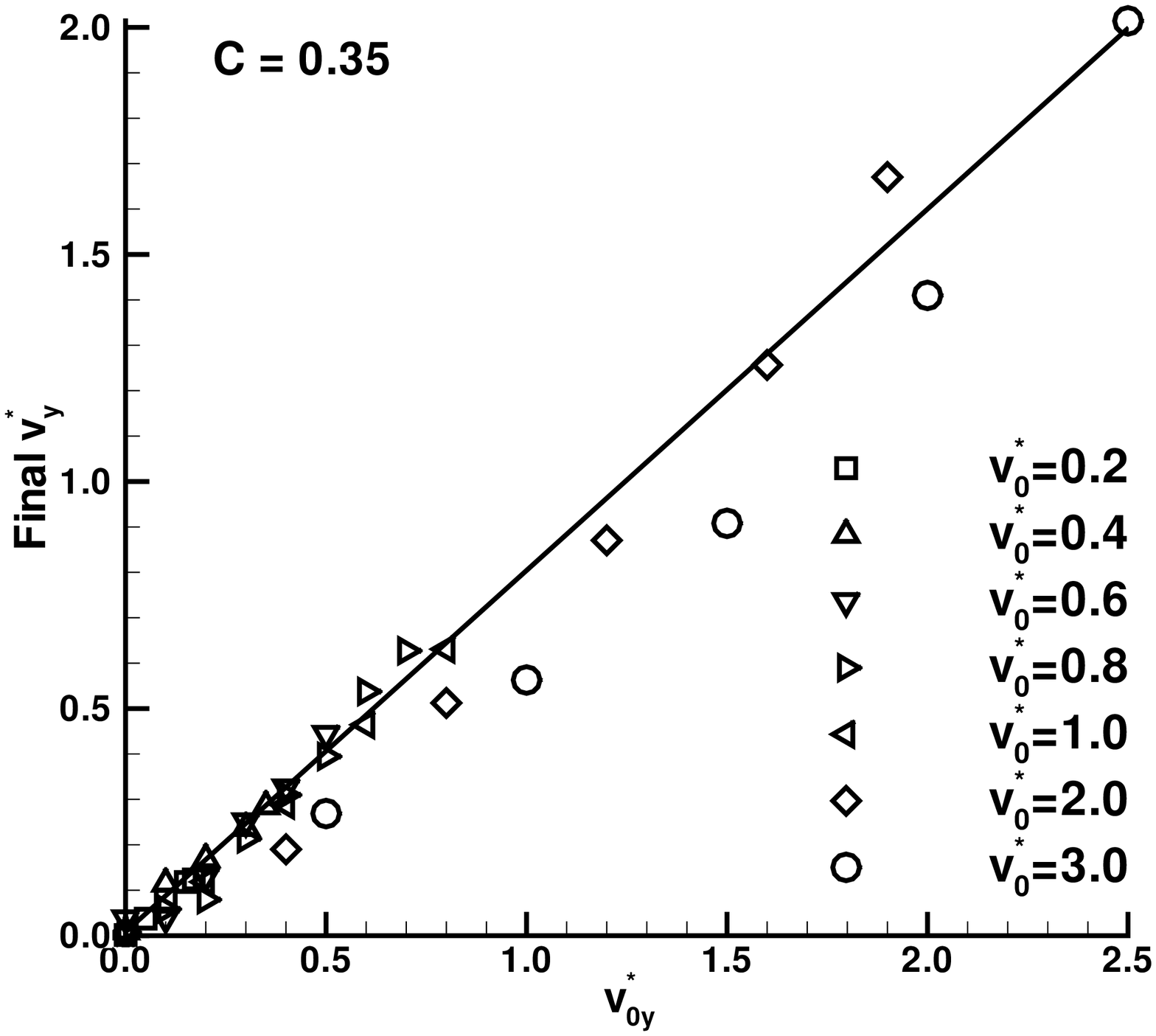}}
{\includegraphics[height=7.5cm]{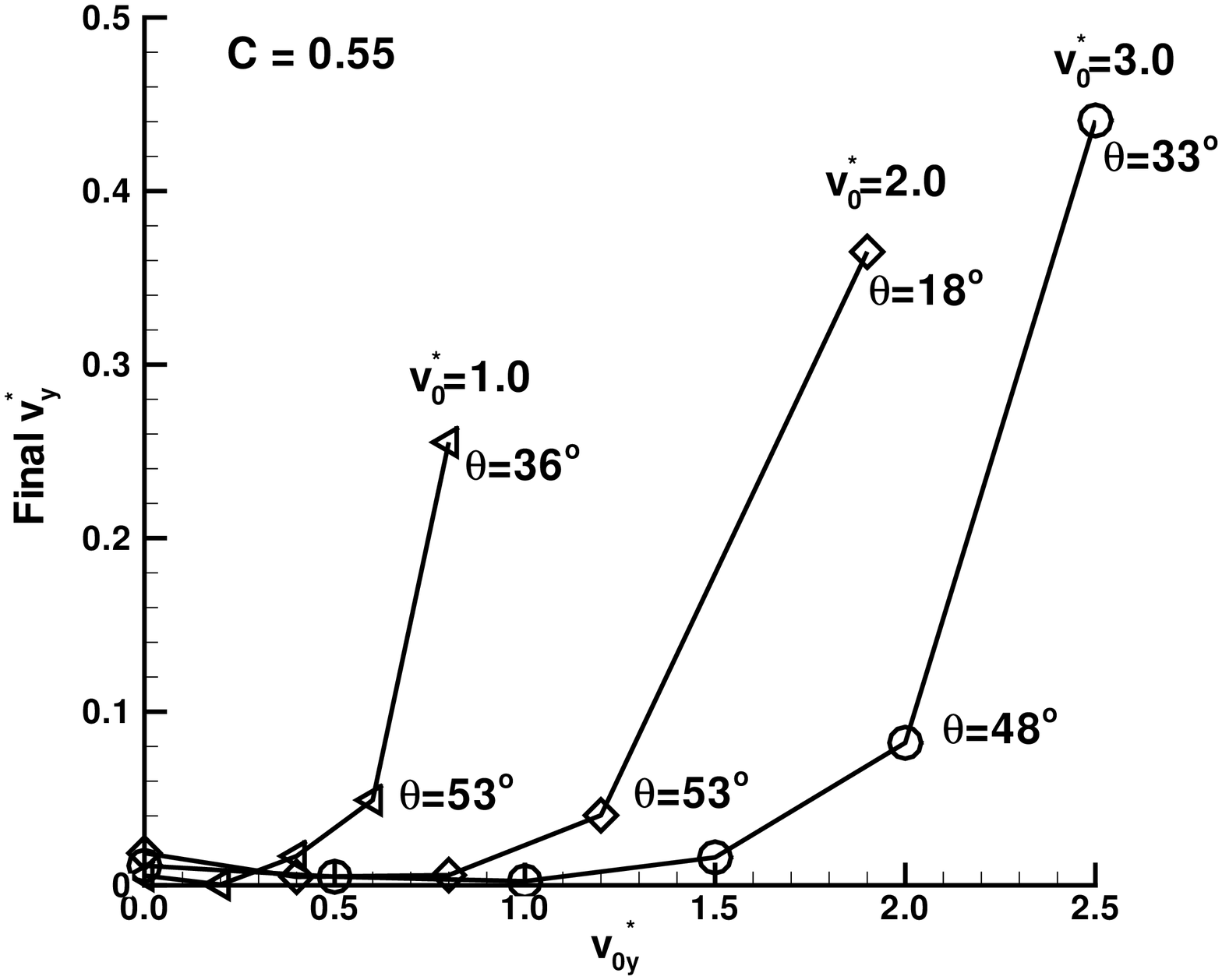}}
\caption{\label{anglesFigure}The final velocity component parallel to the substrate of clusters that are captured by the substrate as a function of the initial velocity component parallel to the substrate, $v^*_{0y}$, for $C=0.35$ (top) and $C=0.55$ (bottom). The clusters slide easily for $C=0.35$ as indicated by the linear dependence of the final velocity on the initial velocity component but for $C=0.55$ there is a threshold incident angle to the surface before sliding takes place.}  
\end{figure}

Figure~\ref{anglesFigure} shows the velocity component {\em parallel} to the surface at the end of the simulation averaged over the clusters that {\em adhere} to the surface at $C=0.35$ (a) and $C=0.55$ (b). At $C=0.35$ it can be seen that this velocity component is approximately conserved during the collision (the slope of the fitting line is 0.8) so that clusters landing and adhering with velocity components parallel to the surface may be characterized as sliding rather than sticking (although these clusters are still counted as adhering in Fig.~\ref{ObliqueStickFigure}). At $C=0.55$ we see a quite different type of behavior, where there seems to be a threshold velocity for sliding that depends on the angle of incidence. Thus for stronger cluster-surface adhesions, clusters striking the surface away from the normal will stick at low velocities and slide at high velocities (see Ref.~\cite{partridge04}).
    
\section{\label{discuss}Discussion}

In this section we compare our simulated results with the experimental data for collisions of Sb, Bi \cite{partridge04} and Cu \cite{reichel06} clusters with SiO$_2$ surfaces. However, as we have not directly simulated a metal/SiO$_2$ system, and the particle sizes here are smaller than those investigated experimentally, we need to be somewhat circumspect in making this comparison. Thus to gain an understanding of how much of the behavior observed is an artefact of the Lennard-Jones model used here, it is useful to draw on knowledge of bulk collisions. In addition, a comparison of our results with bulk collisions is an interesting end in its own right. 

We have investigated a collision regime with typical velocities $v \sim \left(\varepsilon/m\right)^{1/2}$. For Ar, with $\varepsilon = 1.65 \times 10^{-21}$ J/atom, this characteristic velocity is $v \sim 150$ ms$^{-1}$. As noted in the introduction, the collision of Ar clusters with surfaces has been studied previously, both by experiment \cite{chatelet92, Vach94} and by molecular dynamics simulation \cite{Cleveland92}, but at velocities at least an order of magnitude or more larger than those studied here. Similarly using the binding energies for Sb, Bi and Cu, we find characteristic velocities of $510$, $340$ and $780$ ms$^{-1}$ respectively. In Ref.~\cite{reichel06}, it was estimated that their $R=12.5$ nm Sb and Bi clusters had impact velocities of 65 and 45 ms$^{-1}$ respectively, and that their $R=5$ nm Cu clusters had velocities of approximately 230 ms$^{-1}$. Thus, as a fraction of their cohesive energies, the kinetic energies of the clusters in the experimental deposition of Sb, Bi and Cu clusters are an order of magnitude or more below the scales studied here. 

However, from a macropscopic point of view the outcome of a collision (reflection or adhesion) will be determined by the Weber number {\it We}$= E^k_f/E^a_f$ which depends not only on velocity but also cluster size and the cluster-substrate interaction strength. Indeed this is certainly the case in our simulations as discussed in the previous section. Furthermore, our simulations show that the way in which {\it We} depends on these parameters in turn depends on whether the collisions are in the weak or the strong deformation regimes. Again, this is to be expected based on macroscopic considerations \cite{Wu03}. Thus to best compare the experiments and our simulations we must compare Weber numbers, and to do this we need to know whether the collisions in the experimental system are in the weak or strong deformation regime. In what follows, we will focus on estimating the Weber number for $R=12.5$ nm Sb clusters with estimated deposition velocities of 65 ms$^{-1}$.

\subsubsection{The transition between the weak and strong deformation regimes}

Dimensional analysis suggests the onset velocity for strong plastic deformation, $v^{\dagger}$, should scale as $v^{\dagger} \sim \left( Y/\rho \right)^{1/2}$, where $Y$ is the yield stress and $\rho$ is the density of the particle. Indeed, finite element simulations of collisions in bulk systems\cite{Wu03,Wu05} suggest that a transition between elasto-plastic and fully plastic deformations occurs at velocities $v^{\dagger} \sim 0.1 - 0.04 \left( Y/\rho \right)^{1/2}$. In bulk polycrystalline solids $Y$ is typically smaller by a factor of $10^{-3}-10^{-2}$ than $G$, the shear modulus \cite{Kittel86}. For Sb, by assuming that the yield stress is roughly $10^{-3}-10^{-2}$ of the bulk shear modulus, we estimate that $v^{\dagger} \sim 9-30$ ms$^{-1}$. Thus based on the use of bulk material parameters, it seems likely that the Sb collisions in Ref~\cite{partridge04} occurred in a strongly plastic, large deformation regime.

\begin{figure}
\resizebox{\columnwidth}{!}{\includegraphics{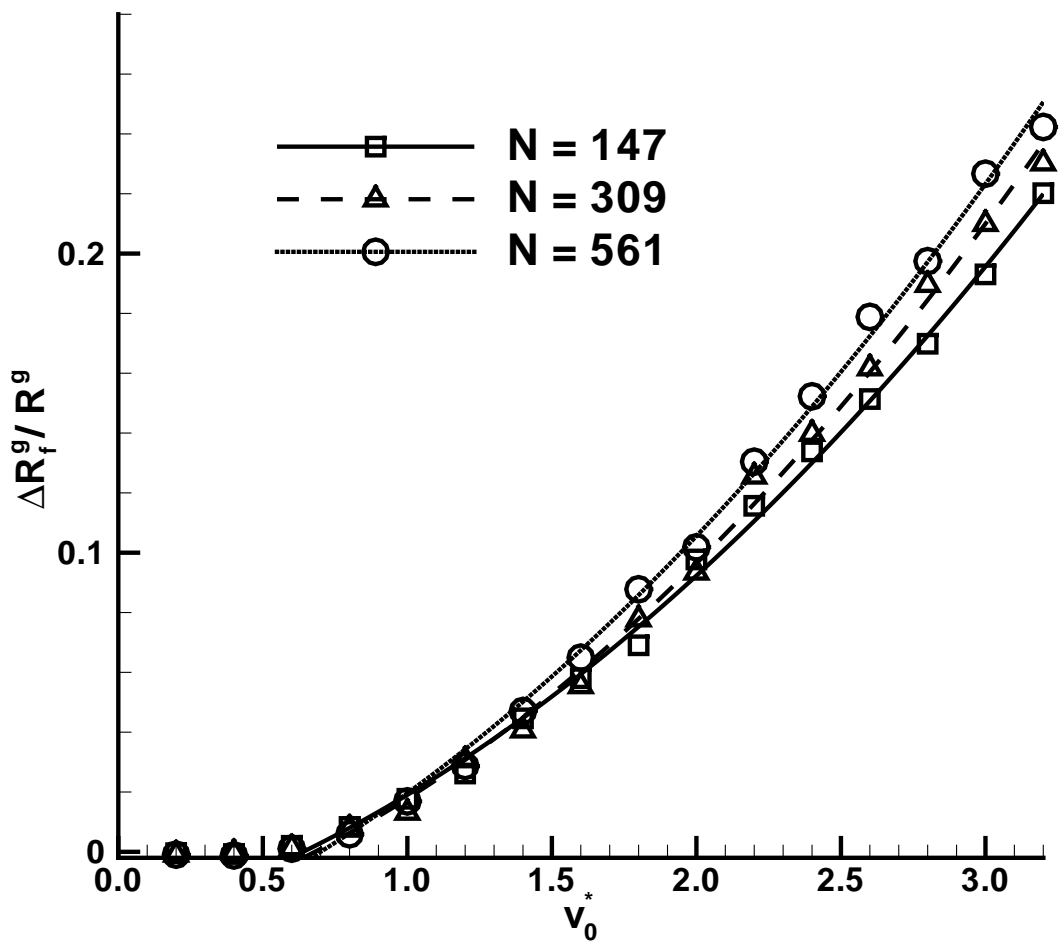}}
\caption{\label{deformFigure} The maximum deformation of the cluster at the moment of peak reflected velocity (the `pull-off'), $v_f^*$, versus impact velocity $v_0^*$ for different size clusters.}  
\end{figure}

To compare this analysis with the behavior of our simulated our system, we need to estimate $Y$ for our clusters. Recalling Fig.~\ref{corFigure}, we note that in the small deformation regime the coefficient of restitution was constant, whereas in the large deformation regime the $e$ exhibits a strong dependence on velocity: $e \sim (v_0^*/v^\dagger)^{-0.6}$. As noted earlier, this is a much stronger dependence on velocity than that given by Hertzian contact mechanics \cite{Johnson87} but is close to the $(v_0^*/v^\dagger)^{-0.5}$ dependence found by the finite-element simulations of strongly plastic collisions \cite{Wu03}. Indeed, the quadratic dependence of the deformation at pull-off on velocity (Fig.~\ref{deformFigure}) is consistent with strong plastic deformation, where the kinetic energy is dissipated largely at the cluster yield stress, $Y$ i.e. the plastic work $\sim Y \Delta (4 \pi R^3/3) \sim 4 \pi Y R^2 \Delta R$ is proportional to the translational kinetic energy $\sim 4 \pi \rho R^3/3 v_0^2 $ so that $\Delta R /R \sim \left(\rho/Y\right) v_0^2$ as is found in Fig.~\ref{deformFigure}. We have found that this relationship is relatively insensitive to both the values of $C$ and the cluster sizes examined here.  

This now allows us to estimate $Y$ for the clusters studied here. Instead of the deformation at pull-off (shown in figure~\ref{deformFigure}), we use the maximum deformation during the collision to estimate $Y$, which is also found to be proportional to the initial particle kinetic energy. If we equate the plastic work to the total kinetic energy, we have $\Delta R /R = A v_0^2$ where $A = \left(\rho/Y\right)$. We can then estimate the constant of proportionality $A$ which then gives $Y \sim 7-12$ $\epsilon/\sigma^3$ for the different cluster sizes. For a Lennard-Jones bulk solid $G_{bulk} \sim 60$ $\epsilon/\sigma^3$ (Ref.~ \cite{quesnel93}) so we have here that $Y \sim 0.1-0.2 \, G_{bulk}$ which is close to the ideal critical shear stress $G/2\pi$. We suspect that there are a several factors that contribute to this relatively large value of $Y$. Firstly, the use of pair potentials such as that in equation (\ref{potential}) is known to lead to simulated materials that are `brittle', in the sense that the onset of plastic deformation only occurs at stresses close to the point of fracture \cite{Holian91}. We also note that many body effects will lead to more anisotropic elastic behaviour \cite{Kum03} but in particles of this size we expect that the anisotropy of the particle structure itself (i.e the icosahedral structure with edges, facets and vertices) will dominate such anisotropic elastic effects. Secondly, the usual mechanisms for plastic deformation that occur in bulk materials, such as the nucleation and propagation of dislocations, are unlikely to operate at the cluster sizes studied here. Indeed, molecular dynamics simulations of metallic nanowires \cite{WAG99} have found yield stresses and Young's moduli considerably above those of the corresponding bulk materials (e.g. $Y_{wire} \sim 10 Y_{bulk}$  with failure occurring via amorphization\cite{WAG99}). Thirdly, we expect that $Y$ will be shear rate dependent (again see Ref.~\cite{WAG99}) and that the very large shear rates in our collisions may lead to higher values of $Y$. Finally, our estimate really represents an upper bound on $Y$ as significant proportion of the incident energy will be be dissipated by the substrate.  

In our simulations, we saw the onset of large plastic deformations at $v^{\dagger} = 0.5 \left(\varepsilon/m\right)^{1/2}$ at all sizes. Based on our estimate of $Y$ above, we find that $v^{\dagger} \sim 0.1 \left( Y/\rho \right)^{1/2}$ which is consistent with Refs.~\cite{Wu03,Wu05}. For Sb, as noted earlier, we estimate that $v^{\dagger} \sim 9-30$ ms$^{-1}$ using bulk material parameters. Nonetheless, if the yield stresses of these Sb particles are significantly larger than $Y_{bulk}$, as we found in our simulated clusters, then $v^{\dagger}$ could be as large as $\sim 90$ ms$^{-1}$. However, in what follows we will proceed with the assumption that the experimental collisions took place in a large deformation, strongly plastic regime as we expect that Sb clusters are likely to considerably more ductile relative to their binding energies than Lennard-Jones clusters.

\subsubsection{The transition between adhesion and reflection}

\begin{figure}
\resizebox{\columnwidth}{!}{\includegraphics{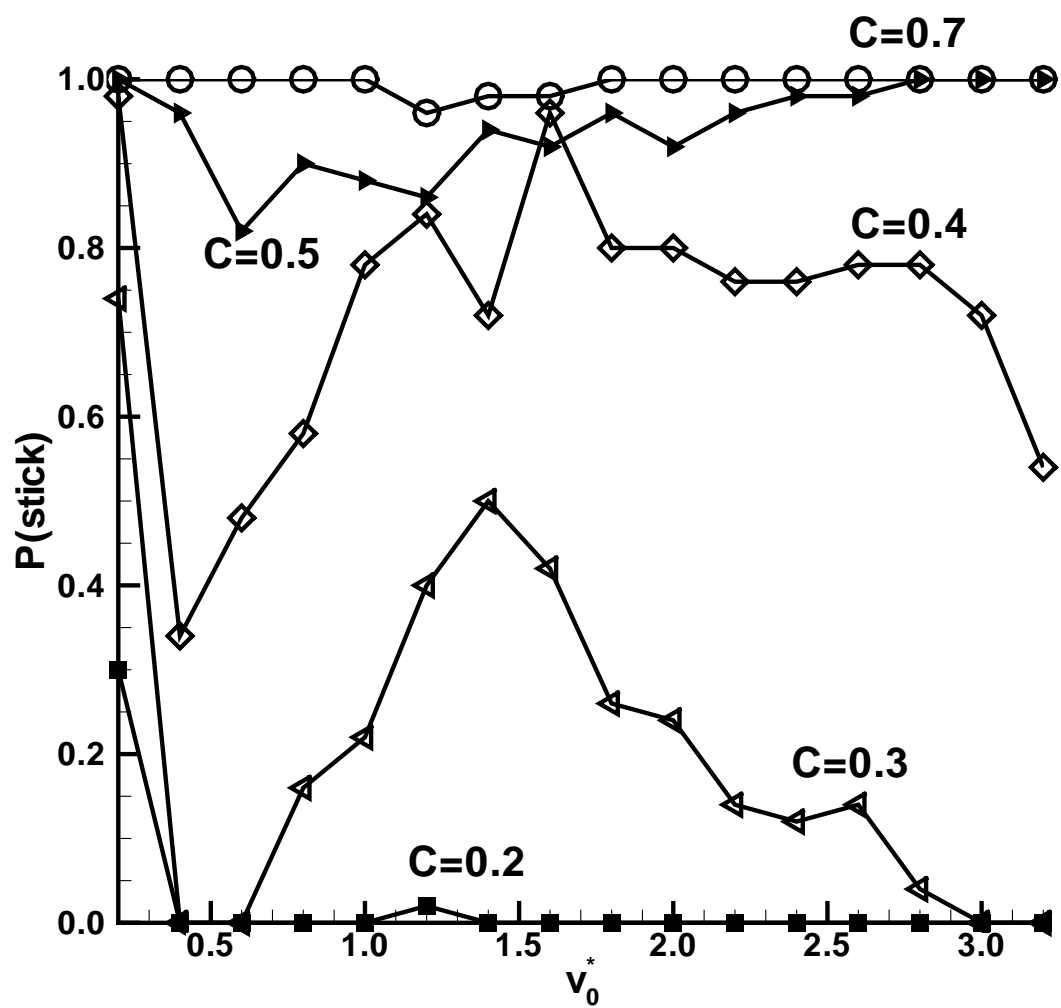}}
\caption{\label{c_Stick} The adhesion probability versus initial velocity ($v^*_{0}$) of 147-atom icosahedral cluster on the flat substrate for different $C$-values.}
\end{figure}

Transitions between adhesion and reflection can evidently occur in either the weak or strong deformation regimes or both (see Fig.~\ref{c_Stick}). 
In the small deformation regime Hertz's contact law for macroscopic bodies \cite{Johnson87} suggests that the contact area should scale as $R^{2/3}$  so that $E^a \sim R^{-7/3}$ (since $E^a$ is the adhesion energy per cluster atom). However, for velocities $v_0^* < 0.5$ in the small deformation regime we find that   $E^a \sim R^{-3/2}$ (see Fig.~\ref{EavsV}(c)). Thus, if we take the coefficient of restitution $e$ to be independent of cluster size, $We \sim v_0^2 R^{3/2}$ in this regime. This is essentially equivalent to the liquid droplet model \cite{hartley58} used in Ref~\cite{partridge04} to explain the transition from adhesion to reflection in their experiments. 

However at impact velocities above $0.5 \left(\varepsilon/m\right)^{1/2}$, in the strong deformation regime, we find that $E^a$ scales as $R^{-1}$. This is consistent with pancaking of the cluster on the surface. It is likely that in the experimental system $E^a$ scales as an inverse power of $R$ with exponent somewhere between the value of $7/3$ which comes from the macroscopic Hertz contact law (which is admittedly only expected to be valid for small deformations) and the value of $1$ seen in the simulations. For the purposes of making a comparison here, we use the latter value obtained from the simulations. Thus in the strong deformation regime, using the macroscopic coefficient of restituion for strong plastic deformation $e \sim (v_0^*/v^{\dagger})^{-0.5}$ \cite{Wu03}, $E^a \sim (v_0^*/v^{\dagger})^{0.5}$ (see section~\ref{sec:pstick}) and $v^{\dagger} \sim 0.1 \left(Y/\rho\right)^{0.5}$, we find that 
\begin{equation}
\label{eqn:weber}
We \sim  \frac{\rho R {v^{\dagger}}^{2}}{\Gamma_a} \left(\frac{v_0^*}{v^{\dagger}}\right)^{0.5} \sim 0.01 \frac{R Y}{\Gamma_a} \left(\frac{v_0^*}{v^{\dagger}}\right)^{0.5}
\end{equation}
where $\Gamma_a$ is the cluster-substrate adhesion energy per unit area. Thus we expect the adhesion-reflection transition to occur at a velocity $v_c$ given by:
\begin{equation}
\label{vcrit}
\frac{v_c}{v^\dagger}\sim 10^2 \left(\frac{\Gamma_a}{R Y} \right)^2
\end{equation}
where $We \sim 1$. The dimensionless groupings on the left and right-hand sides of this expression are quite natural and we would expect these to occur in the description of a reflection-adhesion in any system undergoing plastic deformation. However, as noted above, the scaling with $R$ might vary from system to system, depending on the nature of the adhesive forces involved. Nonetheless, in what follows we will use (\ref{vcrit}) to estimate $v_c$. 

Assuming that the contact angles measured by SEM imaging in Ref.~\cite{partridge04} are in fact the equilibrium contact angles, the adhesion energy can be determined using the surface energy of the metal. The measured contact angle of $\theta \sim 120^o$ for Sb on SiO$_2$ corresponds to an adhesion energy of $\Gamma_a = 0.03$ Jm$^{-2}$, where we have used the surface energy for Sb given in Ref.~\cite{Tyson77}. Thus in a $R=12.5$ nm Sb cluster, if we assume that $Y$ is $10^{-2}$ of the bulk shear modulus, $v_c \sim 6$ ms$^{-1}$ on SiO$_2$. This is consistent with the results reported in Ref.~\cite{partridge04}, where the Sb clusters with deposition velocities of 65 ms$^{-1}$ were reflected from planar surfaces. In the etched V-grooves, the clusters are incident at $35^o$ to the substrate plane, giving a normal velocity component of 35 ms$^{-1}$ which should still result in the clusters being reflected. However, in the V-groove reflected clusters will almost certainly undergo a second impact on the opposite side of the V-groove. This secondary impact is much more likely to lead to adhesion. Thus the experimental results for Sb in Ref.~\cite{partridge04} seem best explained by adhesion following secondary collisions, suggesting that devices can be assembled by depositing clusters well above the transition velocity $v_c$, leading to secondary collisions in a V-groove which occur below $v_c$. 

Finally, we note that there will be {\em at least} two critical velocities for cluster collisions: the velocity $v^\dagger$ which marks the transition between the weak and strong deformation regimes, and the {\em velocities} $v_c$ which give $We = 1$. At these velocities $v_c$, which can occur in the both the weak and strong deformation regimes as seen in Fig.~\ref{weber}, there will be a transition between reflection and adhesion. In our simulated system, this reentrant transition occurs where $v_c^{(1)} < v^\dagger < v_c^{(2)}$ when $C = 0.35$. Fig.~\ref{c_Stick} shows that a similar situation arises for $0.3 < C < 0.4$. In fact this reentrant transition occurs for those values of $C$ where the Weber number is near one at the velocity $v^\dagger$. In other words, if the transition to strong deformation during the collisions occurs at the point where the low deformation collisions have just started to overcome the adhesive forces, particles will begin to adhere once more to the surface due to the increase in adhesion. From equation ($\ref{eqn:weber}$), we see that $We (v^\dagger) \sim 0.01 R Y/ \Gamma_a$ so that for any particular cluster-substrate system, reentrant adhesion is possible in clusters with radii $R \sim 100 \Gamma_a/Y$. In our system, for the cluster sizes we consider, $We(v^\dagger) \sim 1$ when $C \sim 0.3-0.4$. In Sb clusters on the SiO$_2$ surface, this suggests that $We(v^\dagger) \sim 1$ when $R \sim 7$ nm (we note that $We(v^\dagger) = 2$ in the 12.5 nm Sb clusters according to (\ref{eqn:weber})). Thus it may be possible to observe a reentrant transition in $R \sim 7$ nm Sb clusters.

\section{Conclusion}

Using molecular dynamics simulations of Lennard-Jones clusters over a range of velocities, we have observed two collision regimes on weakly attractive substrates in which cluster reflection can occur. At low velocities we find an elastic collision regime, where the cluster progresses from adhesion to reflection as the reflected kinetic energy of the cluster overcomes the adhesion energy. At higher velocities the cluster begins to deform plastically. Initially in this plastic regime the adhesion energy grows faster than the reflected kinetic energy leading to an increase in adhesion probability. However, eventually the reflected kinetic energy grows to dominate the adhesion energy and the adhesion probability decreases once more. As our simulations in this strong deformation regime are consistent with strongly plastic contact mechanics, we propose an expression for the critical velocity for the reflection-adhesion transition in this regime. Using bulk material parameters, we estimate the critical velocity for Sb clusters and find that this is consistent with recent experiments.

\clearpage

\clearpage

\thispagestyle{empty}
\clearpage

\thispagestyle{empty}
\clearpage

\thispagestyle{empty}
\clearpage

\thispagestyle{empty}

\thispagestyle{empty}
\clearpage

\thispagestyle{empty}
\clearpage

\thispagestyle{empty}
\clearpage

\thispagestyle{empty}
\clearpage

\thispagestyle{empty}
\clearpage

\thispagestyle{empty}
\clearpage

\thispagestyle{empty}
\clearpage

\thispagestyle{empty}
\clearpage

\end{document}